\documentclass[fleqn,usenatbib]{mnras}

\usepackage{newtxtext,newtxmath}
\usepackage[T1]{fontenc}
\usepackage{ae,aecompl}

\usepackage{graphicx}	% Including figure files
\usepackage{amsmath}	% Advanced maths commands
\usepackage{amsmath, epsfig,natbib}
\usepackage{multicol}
\usepackage{color,ulem}
\usepackage{newtxtext,newtxmath}
\definecolor{webgreen}{rgb}{0,.5,0}
\definecolor{webbrown}{rgb}{.6,0,0}
\hypersetup{%
   colorlinks=true,hyperfootnotes=false,%
   breaklinks=true,%
   plainpages=false, bookmarksnumbered, bookmarksopen=true,
   bookmarksopenlevel=1,%
   urlcolor=webbrown, linkcolor=blue, citecolor=blue,
   }
\newcommand{\kms}{\mbox{$\>{\rm km\, s^{-1}}$}}
\newcommand{\pc}{\>{\rm pc}}
\newcommand{\kpc}{\mbox{$\>{\rm kpc}$}} 

\newcommand{\Gyr}{\mbox{$\>{\rm Gyr}$}}
\newcommand{\Myr}{\mbox{$\>{\rm Myr}$}}

\newcommand{\Msun}{\>{\rm M_{\odot}}}

\newcommand\degrees{^\circ}
\newcommand\gaia{{\it Gaia}}
\newcommand{\avg}[1]{\mbox{$\left<{#1}\right>$}}

\newcommand{\vb}{\mbox{$V_{\rm breath}$}}
\newcommand{\vbend}{\mbox{$V_{\rm bend}$}}
\newcommand\ie{{i. e.,}}

\title [Spiral driven breathing modes in Milky Way] 
{Age dissection of the vertical breathing motions in \gaia\ DR2: evidence for spiral driving}

\author[S. Ghosh et al.]
	{Soumavo Ghosh$^{1,2}$\thanks{E-mail : ghosh@mpia.de},
        Victor P. Debattista$^{3}$,
        Tigran Khachaturyants$^{3}$\\
        $^{1}$ Max-Planck-Institut f\"{u}r Astronomie, K\"{o}nigstuhl 17, D-69117 Heidelberg, Germany\\
$2$  Inter-University Centre for Astronomy and Astrophysics, Pune 411007, India \\
$^3$Jeremiah Horrocks Institute, University of Central Lancashire, Preston PR1 2HE, UK\\
} 
\date{Accepted 2022 January 14. Received 2022 January 14; in original form 2020 July 20}

\pubyear{2022}

\begin{document}
\label{firstpage}
\pagerange{\pageref{firstpage}--\pageref{lastpage}}
\maketitle

\begin{abstract} 
\gaia\ DR2 has revealed breathing motions in the Milky Way, with stars on both sides of the Galactic mid-plane moving coherently towards or away from it. The generating mechanism of these breathing motions is thought to be spiral density waves. Here we test this hypothesis. Using a {\it self-consistent}, high-resolution simulation with star formation, and which hosts prominent spirals, we first study the signatures of breathing motions excited by spirals. In the model, the breathing motions induced by the spiral structure have an increasing amplitude with distance from the mid-plane, pointing to an internal cause for them. We then show that, at fixed height, the breathing motion amplitude decreases with age. Next, we investigate the signature of the breathing motions in the \gaia\ DR2 dataset. We demonstrate that, at the location with a consistently large breathing motion, the corresponding amplitude increases monotonically with distance from the mid-plane, in agreement with the model. Furthermore, we show that at the same location, the breathing motion amplitude decreases with age, again similar to what we find in the model. This strengthens the case that the observed breathing motions are driven by spiral density waves.

\end{abstract}
%&&&&&&&&&&&&&&&&&&&&&&&&&&&&&&&&&&&&&&

%&&&&&&&&&&&&&&&&&&&&&&&&&&&&&&&&&

\section{Introduction}
%&&&&&&&&&&&&&&&&&&&&&&&&&&&&&&&&&
%
Spirals are common features of disc galaxies in the local Universe \citep[e.g][]{Elmegreenetal2011,Yuetal2018,Savchenkoetal2020}, as well as in high-redshift ($z \sim$ 1.8) disc galaxies \citep[e.g.][]{Elmegreenetal2014,Willetetal2017,Hodgeetal2019}. The nature of spirals has been debated extensively, and, while there is broad agreement that spirals are density waves, a full consensus is yet to emerge \citep[for detailed reviews, see e.g.,][]{DobbsandBaba2014,SellwoodandCarlberg2019}. A wide variety of physical mechanisms have been proposed for exciting spirals, including bars \citep[e.g.,][]{Salo2010,Athanassoulaetal2009,Athanassoulaetal2010,Athanassoula2012,Efthymiopoulosetal2020}, tidal encounters \citep[e.g.,][]{Toomre1972, Dobbsetal2010}, swing amplification of noise \citep{GoldreichLyden65, JulainToomre66, Toomre81}, giant molecular clouds \citep{Donghia2013}, other spirals \citep{Masset1997}, and recurrent groove modes \citep{SellwoodLin1989, SellwoodKahn1991, Sellwood2012, SellwoodandCarlberg2019}. Star formation plays a pivotal role in the persistence of spirals, by cooling the stellar disc and facilitating the generation of fresh spirals \citep{SellwoodCarlberg1984}, while the interstellar gas helps spiral density waves to survive for longer \citep{GhoshJog2015, GhoshJog2016}. The vital dynamical role of spiral arms in transporting angular momentum \citep{LyndenBellKalnajs1972}, as well as in the radial mixing of stars without heating \citep{SellwoodBinney2002, Roskaretal2008, SchonrichBinney2009} has kept the study of the formation and evolution of spirals of continuing interest.

The Milky Way is a barred galaxy \citep{Weinberg1992} which also hosts spiral structure \citep[e.g.,][also see \citet{Valle2005,Valle2008}]{Gerhard2002}. The presence of large-scale, non-zero mean vertical motions of Solar Neighbourhood stars has been reported in various surveys, for example, using main-sequence stars in the Sloan Extension for Galactic Understanding and Exploration (SEGUE) survey \citep{Widrowetal2012}, F-type stars in the Large Sky Area Multi-Object Fibre Spectroscopic Telescope (LAMOST) survey \citep{Carlinetal2013}, and red clump stars in the Radial Velocity Experiment (RAVE) data \citep{Williamsetal2013}. These vertical motions display two distinct types: one in which stars on either side of the mid-plane move coherently in the same direction along the perpendicular to the disc (bending motion) and the other in which stars move coherently in the opposite direction, thereby compressing towards the mid-plane or expanding away from it (breathing motion). In addition, \citet{Widrowetal2012} reported evidence for wave-like north-south asymmetries in the number counts of stars in the Solar Neighbourhood, which were further explored by \citet{YannyGardner2013}. 

The \gaia \ mission \citep{GaiaCollaboration2016, GaiaCollaboration2018} has revolutionised the study of Galactic archaeology by measuring the stellar kinematics of stars in the Solar Neighbourhood with an unprecedented precision. The stellar kinematic study from the second \gaia \ Data Release  (hereafter \gaia\ DR2) revealed rich kinematic substructure in the phase-space distribution of Solar Neighbourhood stars \citep{Antojaetal2018}, as well as the presence of large-scale non-zero vertical motions ($\sim 10 \kms$ in magnitude), and associated bending and breathing motions \citep{Katzetal2018}. Using the RAVE and the Tycho-\gaia \ astrometric solution (TGAS) catalogue, \citet{Carrilloetal2018} showed the presence of both bending and breathing motions in the Solar Neighbourhood. \citet{BennettandBovy2019} revisited the north-south asymmetry in the Solar Neighbourhood stars using the position-velocity measurements from \gaia\ DR2, and confirmed the presence of periodic over- and under-densities of stars in the vertical distribution, regardless of their colours. For an axisymmetric potential, the bulk radial and vertical motions should be zero \citep[e.g.][]{BT08}. Therefore, the question arises what mechanism(s) can induce these non-zero bulk vertical motions in our Galaxy.

Theoretical efforts to understand the cause of vertical motions have explored both external and internal mechanisms. An interaction with a satellite or a dark matter subhalo can excite large-scale coherent vertical bending motions \citep[e.g., see][]{HunterandToomre1969,Araki1985,Mathur1990,Weinberg1991,Gomezetal2013, Widrowetal2014, Donghiaetal2016, Chequersetal2018}.  However bending motions can be produced by purely internal mechanisms, such as the bar \citep{khoperskov2019} or the warp (Khachaturyants et al. {\it submitted}). On the other hand, internal causes such as spiral density waves \citep{Faureetal2014, Debattista2014, Monarietal2016} or the Galactic bar \citep{Monarietal2015} as well as external mechanisms, such as a fly-by encounter with a satellite or passing of a dark matter sub-halo \citep{Widrowetal2014}, have been proposed for exciting the breathing motions. Earlier studies by \citet{Siebertetal2011,Siebertetal2012} showed that spirals and bars can drive a large-scale gradient in the bulk radial velocity in the Milky Way. 
Strong spirals drive large-scale vertical breathing motions ($|\avg{v_z}|~ \sim 5-20 \kms$)  as first shown by \citet{Faureetal2014}. Their semi-analytic models assumed a thin spiral structure with a small vertical extent. 
Their analytic treatment found that the breathing motions vary with distance from the mid-plane as $\tanh(z/z_0) \mathrm{sech}^2(z/z_0)$, where $z_0$ is the scale-height of the spiral. This peaks at $\sim 0.7 z_0$.
 At larger heights, this decreases as $\exp(-2z/z_0)$ \citep[see Eqs~10 \& 12 in][]{Faureetal2014}. The self-consistent simulations of \citet{Debattista2014}, instead, produced vertically extended spirals, which resulted in vertical motions that increase with distance from the mid-plane.  
The relative sense of the bulk motions, whether compressing or expanding changes across the corotation resonance (hereafter CR) of the spiral. Inside the CR, the vertical motions are compressive behind the peak of the spiral and expanding ahead of the spiral's peak. Outside the CR, the compressive and expanding breathing motions reverse their sense of occurrence \citep{Faureetal2014, Debattista2014}.

In this paper, we continue to explore the possibility that the breathing motions observed in \gaia\ DR2 are caused by the Milky Way's spiral structure. Using a high-resolution, self-consistent, star-forming simulation of a Milky Way-like  galaxy, we study the characteristics of the breathing motions driven by spiral structure. Previous studies of breathing motions either used perturbation theory \citep{Monarietal2016} and test particle simulations \citep{Faureetal2014} or used self-consistent simulations but lacked the high mass resolution necessary to sub-sample the stellar populations \citep{Debattista2014}. The novelty of our paper, aside from comparing directly with the \gaia\ DR2 data, lies in the fact that our high-resolution model allows us to explore how different populations participate in the breathing motions. 

The paper is organised as follows: Section~\ref{sec:numerical_details} provides the details of the simulation, including quantifying the spiral structure present at our chosen snapshot, which, for simplicity, we refer to as `the model'. Section~\ref{sec:vert_velocity_nbodySPH} presents the breathing motions of the model.  Section~\ref{sec:novel_technique} provides the details of a novel technique to detect/identify breathing motions from the variation of the bulk vertical motions. Section~\ref{sec:gaia_comparison} compares the vertical kinematic signatures of Milky Way stars with the trends obtained in the model. Section~\ref{sec:physical_explanation} discusses the {\it unique} fingerprint of spiral-driven vertical breathing motions, while Section~\ref{sec:discussion} summarises our main conclusions.

\section{Star-forming simulation}
\label{sec:numerical_details}
%&&&&&&&&&&&&&&&&&&&&&&&&&&&&&&&&&&&&&&&

We use an $N$-body$+$smooth particle hydrodynamics (SPH) simulation with gas and star formation to motivate some of the analysis of the \gaia\ DR2 data. This is a higher mass resolution version of the models described in our previous papers \citep[e.g.][]{Roskaretal2008, Loebmanetal2016}.  This model starts with a gas corona in pressure equilibrium with a co-spatial Navarro-Frenk-White \citep{nfw} dark matter halo which constitutes $90\%$ of the mass. The dark matter halo has a virial radius $r_{200} \simeq 200 \kpc$ and a virial mass $M_{200} = 10^{12} \Msun$. Gas velocities are given a spin of $\lambda = 0.065$ \citep{Bullocketal2001}, with specific angular momentum proportional to radius. Both the gas corona and the dark matter halo consist of $5 \times 10^6$ particles; gas particles have softening $\epsilon = 50 \pc$, while that of dark matter particles is $\epsilon = 100 \pc$. All stars form out of cooling gas, inheriting the softening of the parent gas particle.  We evolve the simulation for $13 \Gyr$ with {\sc gasoline} \citep{Wadsleyetal2004, Wadsleyetal2017}. As gas cools it settles into a disc; once the gas density exceeds 0.1 cm$^{-3}$ and the temperature drops below 15,000 K, star formation and supernova blast wave feedback commence \citep{Stinsonetal2006}. Supernovae feedback couples $40\%$ of the $10^{51}$ erg per supernova to the interstellar medium as thermal energy. Gas mixing uses turbulent diffusion as described by \citet{Shenetal2010}.

We use a base timestep of $\Delta t=5\Myr$ with timesteps refined such that $\delta t = \Delta t/2^n < \eta\sqrt{\epsilon/a_g}$, where we set the refinement parameter $\eta = 0.175$. We set the opening angle of the tree-code gravity calculation to $\theta = 0.7$. Gas particle timesteps also satisfy the condition $\delta t_{gas} = \eta_{courant}h/[(1 + \alpha)c + \beta \mu_{max}]$, where $\eta_{courant} = 0.4$, $h$ is the SPH smoothing length set over the nearest 32 particles, $\alpha$ and $\beta$ are the linear and quadratic viscosity coefficients and $\mu_{max}$ is described in \citet{Wadsleyetal2004}. Star particles represent single stellar populations with a Miller-Scalo initial mass function.

\subsection{Spiral structure in the model}
\label{ssec:quantify_spiral}
%&&&&&&&&&&&&&&&&&&&&&&&&&&&&&&&&&&&&&&&

We use a snapshot from this model at $t=12\Gyr$.  We choose this snapshot strictly because it allows a good division of its stellar populations, not because of any special properties of the model at this time.  While the simulation we use has not been published before, the general evolution of the spiral structure in this class of models is explored in \citet{Roskaretal2012}.

\begin{figure*}
    \includegraphics[width=\linewidth]{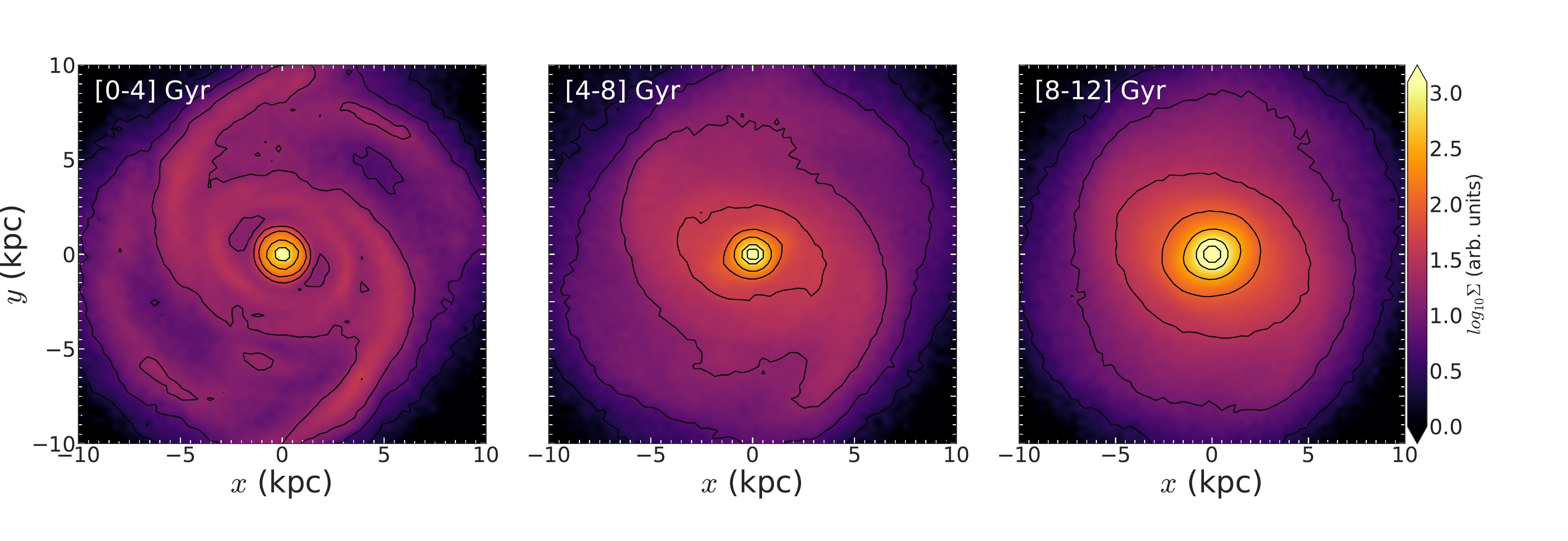}
    \medskip
    \includegraphics[width=\linewidth]{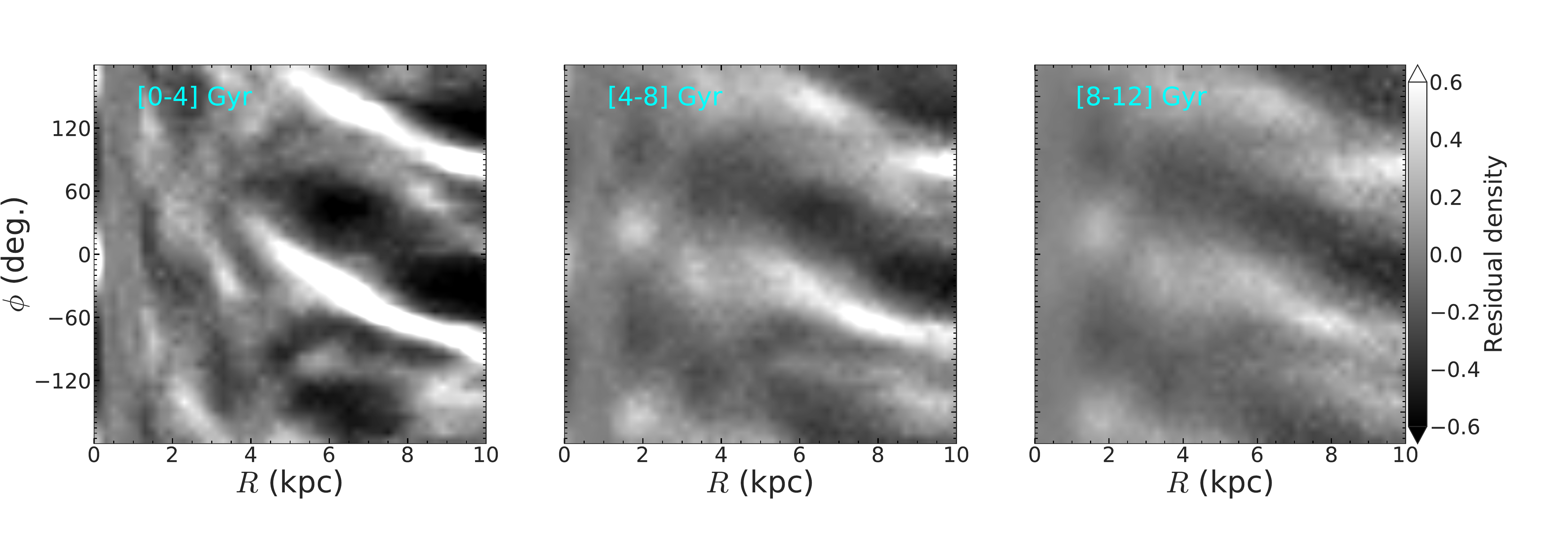}
\caption{Star-forming model: distribution of stellar surface density ({\it top panels}), and the residual surface density ({\it bottom panels}), calculated using Eqn.~\ref{eq:resi_density} at $t = 12 \Gyr$ for (left to right) young (ages $=0-4 \Gyr$), intermediate (ages $=4-8 \Gyr$), and old (ages $= 8-12 \Gyr$) populations. Contours of constant surface density (black solid lines) are overlaid on the surface density maps. Unambiguous spiral structure is present in the young and intermediate age populations, and the residuals reveal that the old population also supports the spiral. The sense of rotation is towards increasing $\phi$.}
\label{fig:density_residualmaps_nbodysph}
\end{figure*}

In order to facilitate comparison with the Milky Way, we quantify the spiral amplitude in the model to show that the spirals in the model are not unreasonably strong. We also quantify the strength of the spiral as a function of stellar age. We divide the stars into three broad age bins: a young stellar population (with ages between 0-4~\Gyr), an intermediate-age stellar population (with ages between 4-8~\Gyr), and an old stellar population (with ages between 8-12~\Gyr). The top panels of Fig.~\ref{fig:density_residualmaps_nbodysph} show the surface density of stars in the face-on, $(x,y)$, view for the three age populations. The bottom panels show the residual surface density, $\tilde \Sigma (R, \phi)$, calculated as:
\begin{equation}
\tilde\Sigma (R, \phi)=\frac{\Sigma(R, \phi)- \Sigma_{\rm avg}(R)}{\Sigma_{\rm avg}(R)}\,,
\label{eq:resi_density}
\end{equation}
where $\Sigma_{\rm avg}(R)$ is the average density at radius $R$, and the averaging is carried out over the whole azimuthal range.

The density plots of Fig.~\ref{fig:density_residualmaps_nbodysph} show that the youngest population exhibits a very prominent two-armed spiral structure, which is also present in the intermediate-age population, but is indistinct in the old population. However the residual density plots show that the $m=2$ spiral is present even in the old population, albeit weaker.  This therefore is a genuine density wave which propagates in all stellar populations.  The estimated pitch angle, $\gamma$, of the spiral in this snapshot is $\sim 33^{\circ}$. In comparison, the Milky Way’s spirals are tighter, having a pitch angle $\gamma \sim 10^{\circ}$ \citep{Siebertetal2012}.

\begin{figure}
\centering
     \includegraphics[width=1.\linewidth]{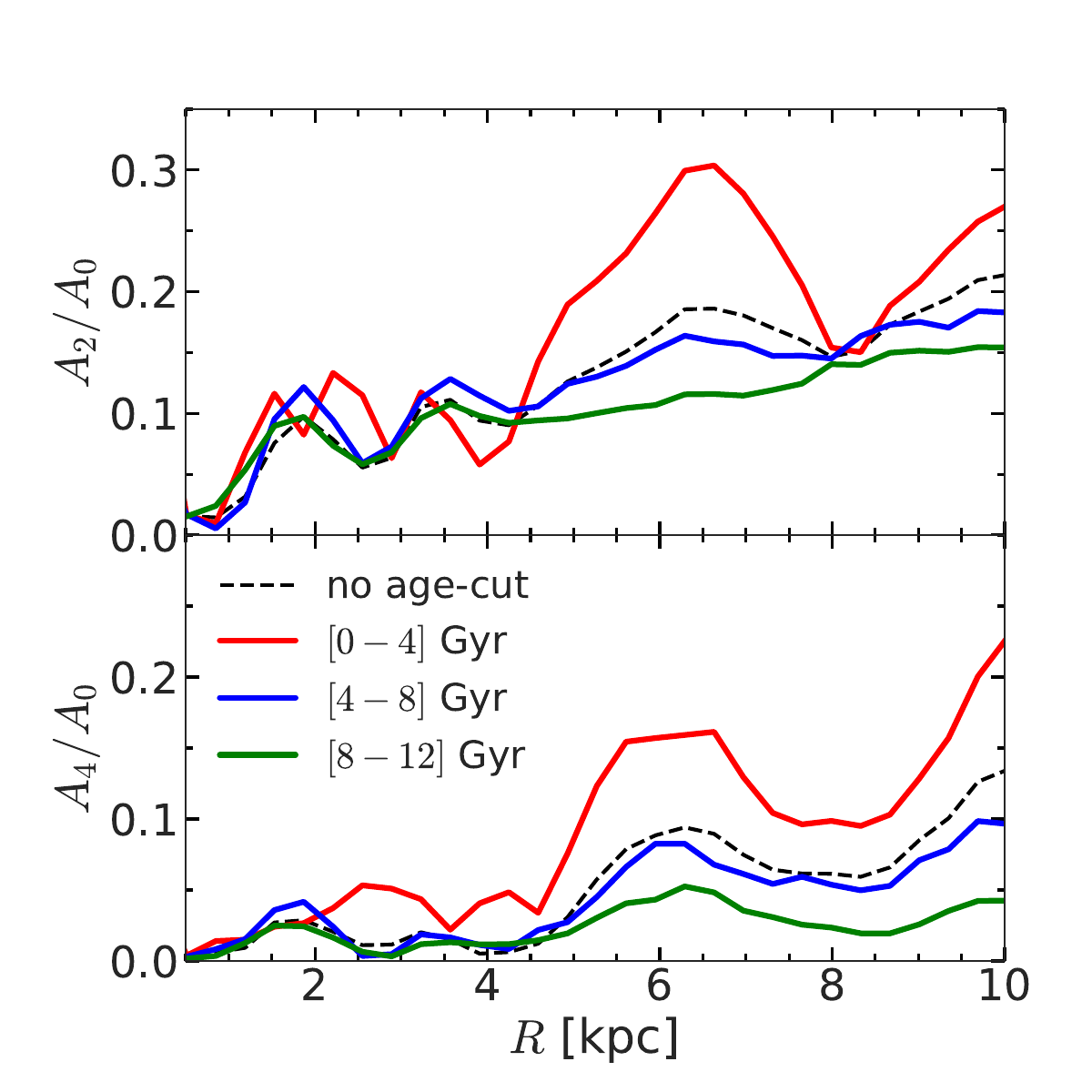}
\caption{Star-forming model: radial profile of $m=2$ (top panel) and $m=4$ (bottom panel) Fourier moments of the surface density for the three stellar populations at $t = 12 \Gyr$.  The nearly zero $A_4$ in the old population indicates that the spiral in these stars is more gently varying with azimuth than in younger populations.  Note the change of vertical axis scale between the two panels. The black dashed lines denote the Fourier moments for all stars.}
\label{fig:denfourier_t12_agecut}
\end{figure}

To quantify the strength of the spirals, we measure the radial profiles of the Fourier moments of the surface density, defined as:
\begin{equation}
A_m/A_0 (R)= \frac{\sum_i m_i e^{im\phi_i}}{\sum_i m_i}
\label{eq:fourier_calc}
\end{equation}
where $A_m$ is the coefficient of the $m^{th}$ Fourier moment of the density distribution, $m_i$ is the mass of the $i^{th}$ particle and $\phi_i$ is its cylindrical angle. The peak of the Fourier $m=2$ amplitude, $(A_2/A_0)_{\rm max} \simeq 0.19$ for our model. In comparison, \citet{Rix1995} found that, for their sample of galaxies with $m=2$ spirals, the peak value of $A_2/A_0$ varies in the range 0.15 to 0.6, with $\sim 28\%$ of their sample having $(A_2/A_0)_{\rm max} \leq 0.2$.  The spiral structure in the model therefore is sufficiently realistic to provide a basis for comparing the kinematic signature of the spiral in the model with that in the Milky Way.
The amplitudes for the $m=2$ and $m=4$ Fourier harmonics for all three stellar populations at $t= 12~\Gyr$ are shown in Fig.~\ref{fig:denfourier_t12_agecut}. 
The spiral has a significant $m=2$ amplitude  at $4 \leq R/\kpc \leq 8$.  While the amplitude decreases with stellar age, all stellar populations take part in the spiral structure, as opposed to {\it material spirals} present in only the young stellar populations \citep[for detailed discussion, see e.g.][]{BT08,DobbsandBaba2014}.  The $m=4$ amplitude is likewise strongest in the young population, and is almost zero for the old population.  Thus the spiral in the old population varies more gently in azimuth while that in the young population is more sharply delineated, as can be seen in Fig. \ref{fig:density_residualmaps_nbodysph}.

\begin{figure}
\centering
     \includegraphics[width=1.05\linewidth]{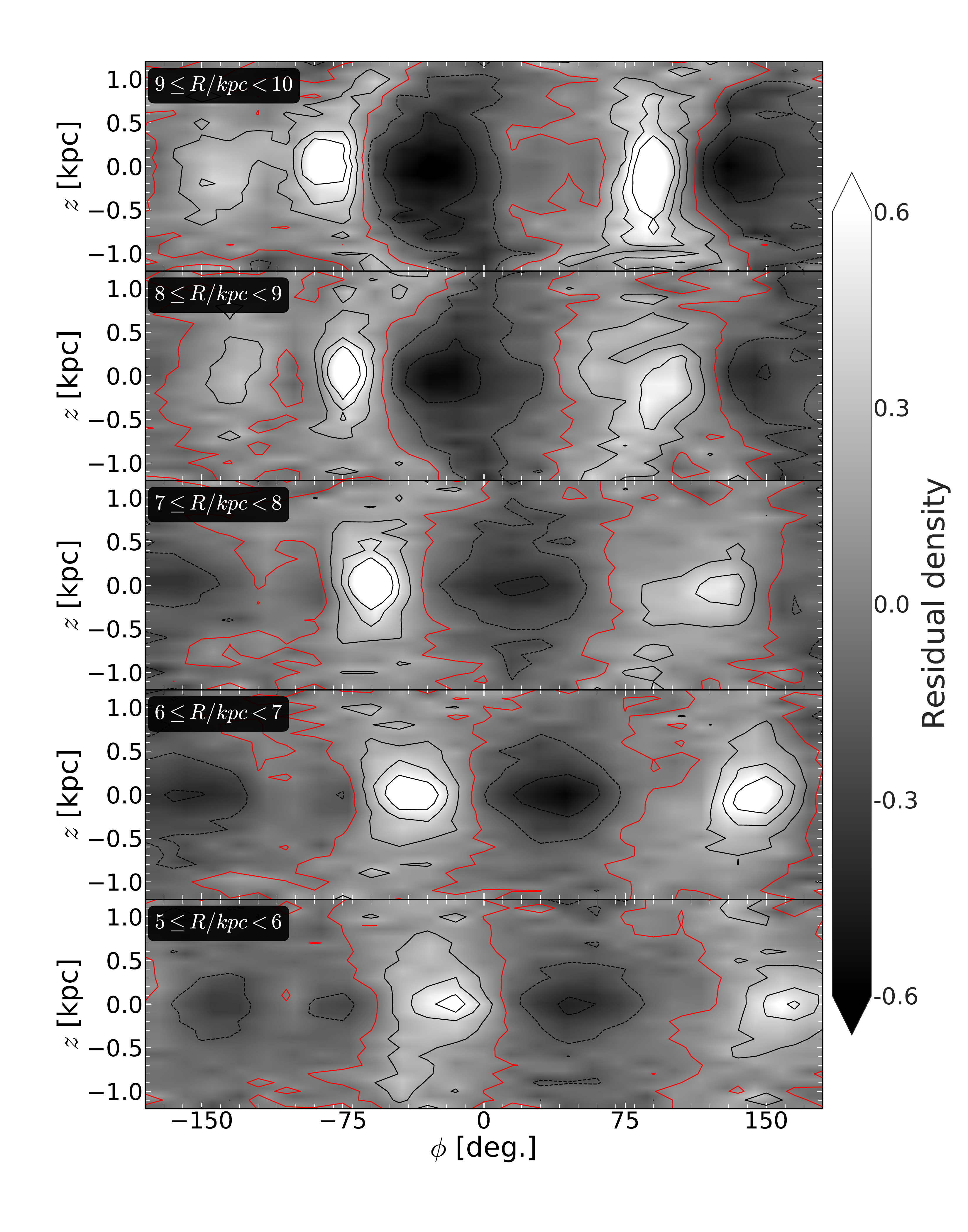}
\caption{Star-forming model: residual surface density, $\delta \Sigma (\phi, z; \Delta R)$, calculated using Eqn.~\ref{eq:resi_density_phiz}, shown in the $(\phi, z)$~plane at five different radial annuli of width $\Delta R = 1 \kpc$. The contours indicate constant residual surface density, with the red contours denoting $\delta \Sigma (\phi, z; \Delta R) = 0$. The sense of rotation is towards increasing $\phi$.}
\label{fig:residualmap_phiz}
\end{figure}

We also calculate the residual density distribution $\delta\Sigma (\phi, z)$ within a radial annulus of width $\Delta R$ as
\begin{equation}
\delta \Sigma (\phi, z; \Delta R) = \frac{\Sigma(\phi, z; \Delta R)- \Sigma_{\rm avg}(z; \Delta R)}{\Sigma_{\rm avg}(z; \Delta R)}\,,
\label{eq:resi_density_phiz}
\end{equation}
where $\Sigma_{\rm avg}(z; \Delta R)$ is the average density, and the averaging is carried out over the whole azimuthal range within the radial annulus of radial width $\Delta R$, at a fixed height $z$. Fig.~\ref{fig:residualmap_phiz} shows the resulting residual maps in the $(\phi, z)$~plane at different  annuli of $\Delta R = 1 \kpc$. At each  annulus, the peak density in the mid-plane ($z = 0$) is ahead (relative to the sense of rotation) of the peak density at larger heights from the mid-plane. Also, the density distribution at fixed $R$ and $z$ is azimuthally asymmetric relative to the peak density, in agreement with the findings of  \citet{Debattista2014}. The spiral extends to the full vertical extent, rather than being limited to close to the mid-plane.  This behaviour was also found in pure $N$-body simulations by \citet{Debattista2014}, and will be important for our subsequent analysis.

\section{Breathing motions induced by spirals }
\label{sec:vert_velocity_nbodySPH}
%&&&&&&&&&&&&&&&&&&&&&&&&&&&&&&&&&&&&&&&&&&&&

We start by studying the breathing motions induced by the spirals in the star-forming model in order to explore how the breathing motions vary between different age populations. The self-consistent simulation of \citet{Debattista2014} showed that the amplitudes of the breathing motion increases with height from the mid-plane; here we extend that earlier work but exploring how the amplitudes change as a function of both height and stellar age.

We define the breathing velocity, $V_{\rm breath} $, as \citep{Debattista2014, Widrowetal2014, Katzetal2018}:
\begin{equation}
\vb(x, y) = \frac{1}{2}\left[\avg{v_z(x,y, \Delta z)} - \avg{v_z(x,y, -\Delta z)}\right]\,,
\label{eq:vbreath}
\end{equation}
where $\avg{v_z(x,y,\Delta z)}$ is the mean vertical velocity at position $(x,y)$ in the galactocentric cartesian coordinate system, averaged over a horizontal layer of thickness $\Delta z$ \citep[for details see][]{Katzetal2018}. We choose $\Delta z = 400 \pc$, and calculate $\vb $ in three such vertical layers: $|z| = [0, 400] \pc$, $|z| = [400, 800] \pc$ and $|z| = [800, 1200] \pc$. In this definition of \vb, the average in a particular layer is carried out before the difference between the two sides is taken. In the observational data this helps reduce the effect of any selection function differences between the two sides of the disc. This also removes the trivial effect of bending waves, which would otherwise give rise to a spurious signature by displacing stars to one side of the mid-plane. Eq.~\ref{eq:vbreath} implies that when $\vb > 0$ then stars are coherently moving away from the mid-plane (expanding breathing motion), while $\vb < 0$ when stars are moving towards the mid-plane (compressing breathing motion). 
For the sake of comparison, we also consider the bending velocity:
\begin{equation}
V_{\rm bend}(x, y) = \frac{1}{2}\left[\avg{v_z(x,y, \Delta z)} + \avg{v_z(x,y, -\Delta z)}\right].
\label{eq:vbend}
\end{equation}
The star-forming model, which is not externally perturbed, lacks substantial coherent bending waves, so the model's bending velocities are not shown here.

\subsection{Bulk vertical motions of the different age populations}
\label{subsec:meridional_plane_nbodySPH}
%&&&&&&&&&&&&&&&&&&&&&&&&&&&&&&&&&&&&&&&&&&&&&&&&&&&&&&&&&&&&&&&&&&&&&&&&

\begin{figure*}
    \includegraphics[width=0.95\linewidth]{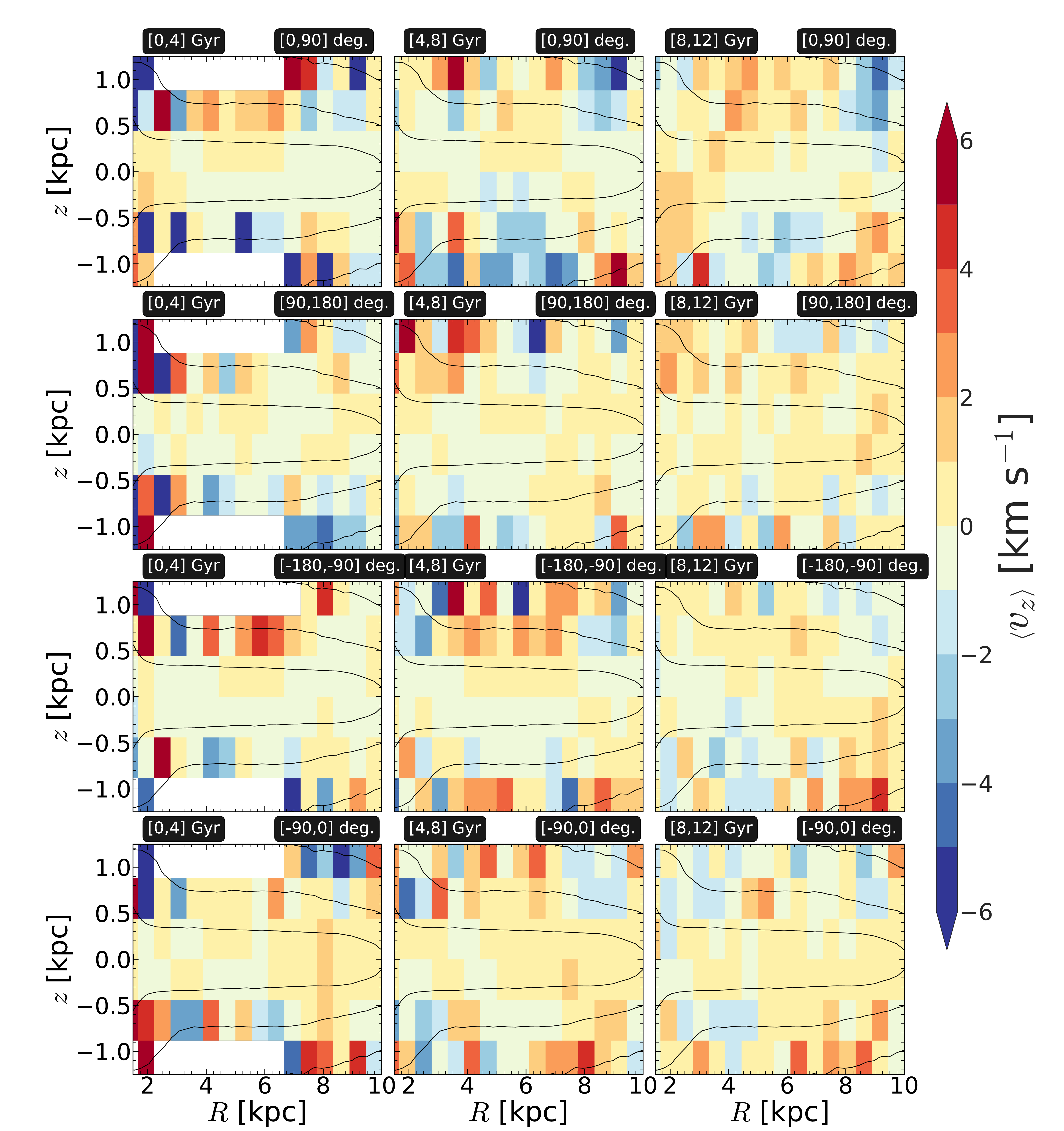}    
 
\caption{Star-forming model: distribution of bulk vertical velocity, $\avg{v_z}$, in the meridional, $(R,z)$, plane, for the three different stellar populations at $t = 12~\Gyr$. {\it Left to right}: stars age $=0-4~\Gyr$ (young), $=4-8~\Gyr$ (intermediate), and $8-12~\Gyr$ (old).  {\it Top to bottom}: stars at different azimuthal wedges are shown. The age bins and the azimuthal wedges are listed above each sub-plot. Contours of the {\it total} density are overlaid in each panel.}
\label{fig:meanvz_Rzmap_nbodysph}
\end{figure*} 

We examine the star-forming model's vertical motions as a function of stellar age. The distribution of the bulk vertical motions in the meridional, $(R,z)$, plane, calculated at four different azimuthal wedges, for the three stellar populations at $t = 12~\Gyr$, are shown in Fig.~\ref{fig:meanvz_Rzmap_nbodysph}. Large-scale, non-zero vertical motions are present in all three populations. The absolute values of the mean vertical velocity, $|\avg{v_z}|$, are small, $\sim 5\kms$, but non-zero, increasing with height from the mid-plane. There are not enough stellar particles in the young population at large heights to measure a reliable mean vertical velocity for them there. The intermediate-age population exhibits a large $|\avg{v_z}|$ at larger heights; $|\avg{v_z}|$ gets somewhat weaker for the old stellar population. All three populations are coherently moving away from, or towards, the disc mid-plane indicating breathing motions predominate in all stellar populations. There are also bending motions present, but these are weaker compared to the breathing  motions. These bending motions are studied in detail elsewhere (Khachaturyants et al. {\it submitted}). All three age populations exhibit prominent bulk vertical breathing motions in the $(R,z)$ plane. However, the azimuthal wedges used in Fig.~\ref{fig:meanvz_Rzmap_nbodysph} are large, therefore the coincidence of strong breathing motion with the spiral structure is not readily apparent in this figure. This is further demonstrated in Section~\ref{sec:novel_technique}.

\begin{figure}
\centering
     \includegraphics[width=1.05\linewidth]{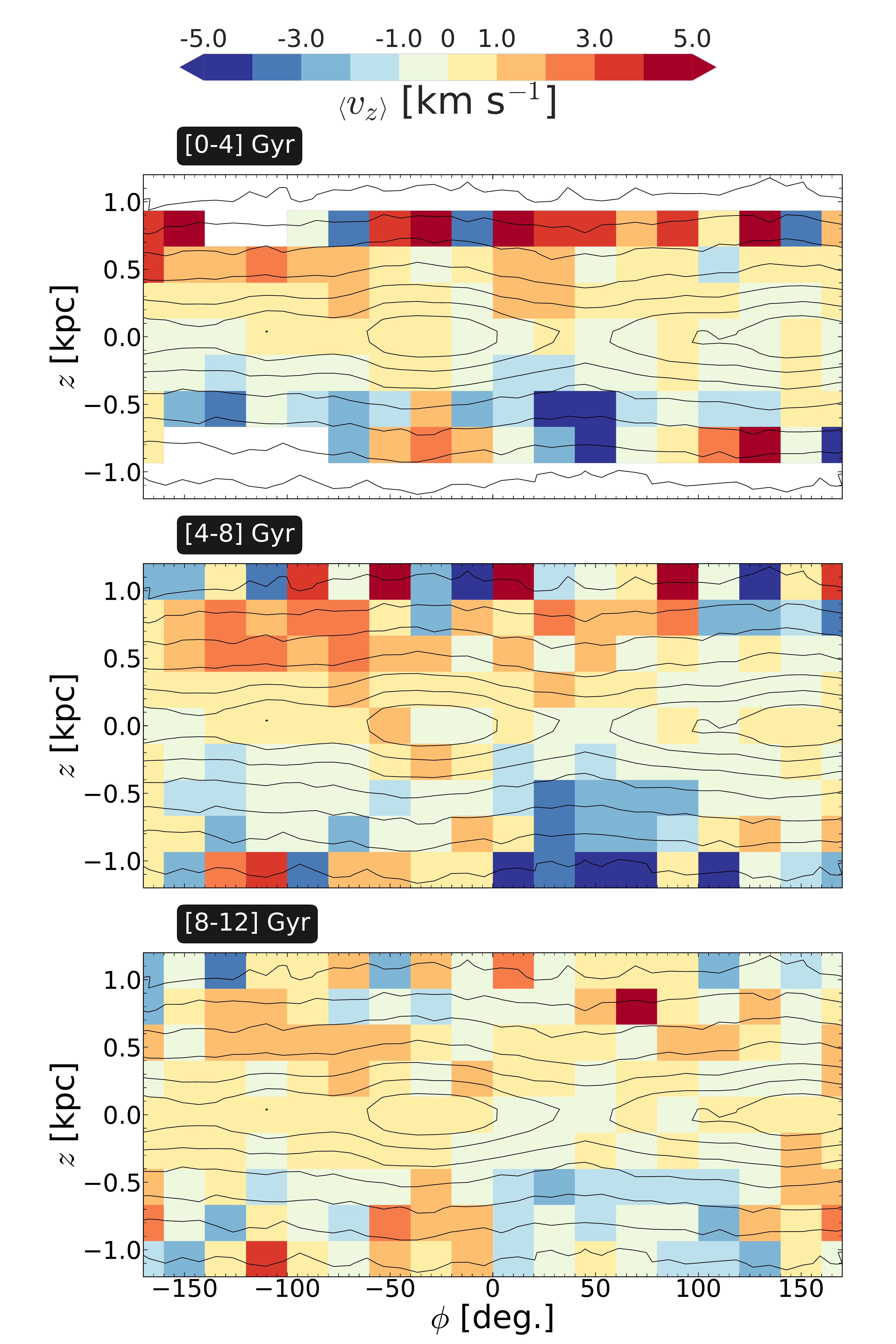}
\caption{Star-forming model: distribution of $\avg{v_z}$ in the $(\phi,z)$~plane, for stars in the radial annulus $5 \le R/\kpc \le 7$ at $t = 12 \Gyr$. {\it Top:} the young stellar population, {\it middle:} the intermediate-age stellar population, and {\it bottom:} the old stellar population. The presence of large-scale, non-zero bulk vertical velocities for stars in all three populations is visible, and are coherent across age.  The vertical motions are smallest in the old population. The sense of rotation is towards increasing $\phi$.
}
    \label{fig:meanvz_phizmap_agecut_nbodySPH}
\end{figure}

We then select stars in the radial annulus $5 \le R/\kpc \le 7$,  calculate \avg{v_z}\ in the $(\phi,z)$~plane for the different stellar populations, and plot these in Fig.~\ref{fig:meanvz_phizmap_agecut_nbodySPH}.  Non-zero bulk vertical motions are evident in all the stellar populations, although the young population particle numbers are not large enough at the uppermost slice to measure a meaningful $\avg {v_z}$. These results indicate breathing motions are present in this annulus.  The vertical motions switch from compressive to expanding at $\phi \simeq 0\degrees$, which Fig.~\ref{fig:density_residualmaps_nbodysph} shows is the location of the peak density in the youngest population.  The amplitude of the vertical motions decreases with age, with the old population having velocities roughly half those of the intermediate population.

\subsection{Breathing motions of the different age populations}
\label{subsec:breathing_velocity_nbodySPH}
%&&&&&&&&&&&&&&&&&&&&&&&&&&&&&&&&&&&&&&&&&&&&&&&&&&&&&&&&&&&&&&&&&&&&&&&&

\begin{figure*}
    \includegraphics[width=1.05\linewidth]{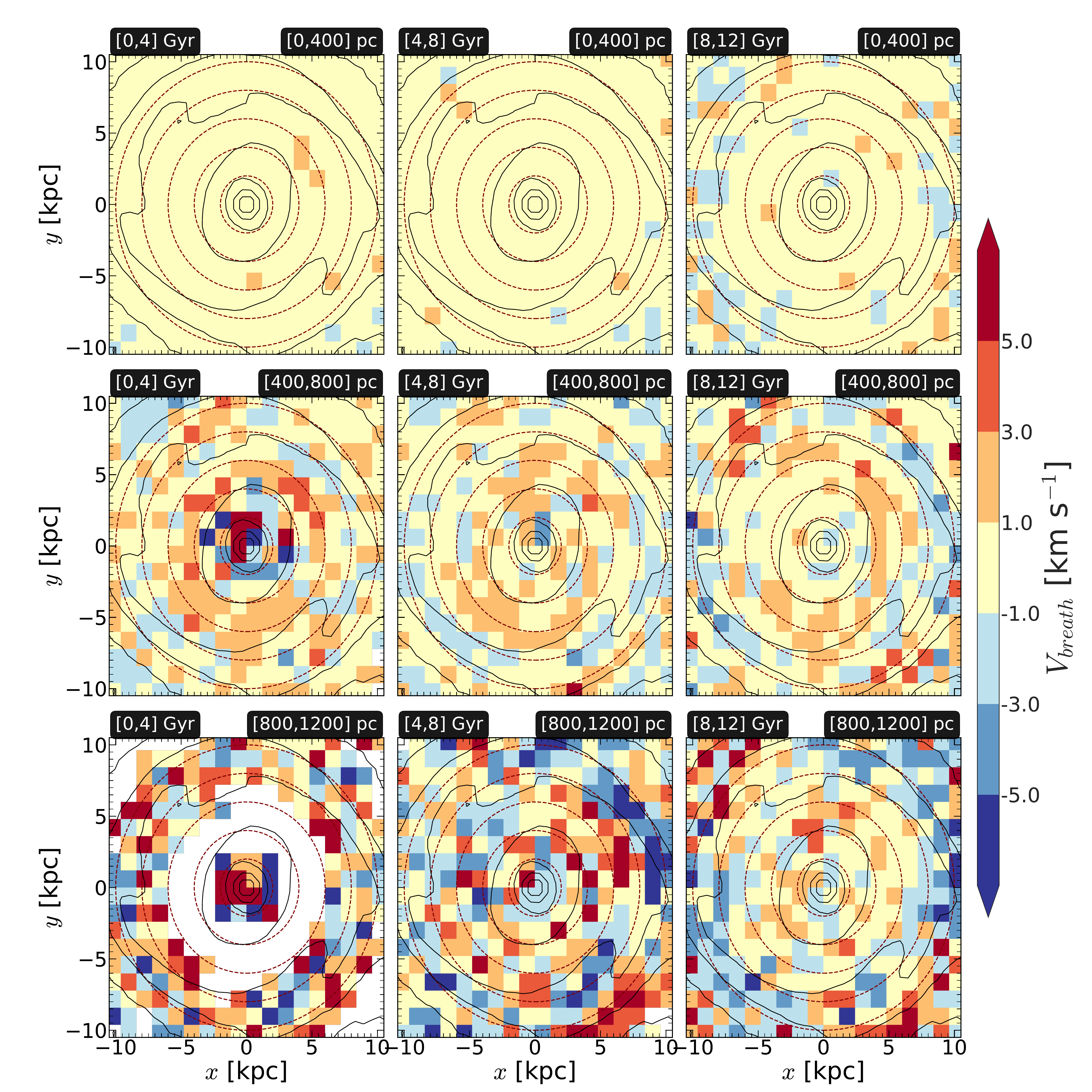}
   
\caption{Star-forming model: breathing velocity, \vb, for stars of different ages at different vertical distances from the mid-plane, shown at $t = 12 \Gyr$. From {\it left to right}: stars with ages $0-4 \Gyr$ (young), $4-8 \Gyr$ (intermediate), and $8-12 \Gyr$ (old), respectively, whereas from {\it top to bottom}: stars at $|z| = [0, 400]~\pc$, at $|z| = [400, 800]~\pc$, and at $|z| = [800, 1200]~\pc$, respectively. Contours of {\it total} surface density are overlaid in each panel. The dashed circles (in maroon) indicate galactic radii ranging from $2 \kpc$ to $10 \kpc$, in $2 \kpc$ steps.} 
\label{fig:breath_xymap_nbodysph}
\end{figure*}

\begin{figure*}
    \includegraphics[width=1.05\linewidth]{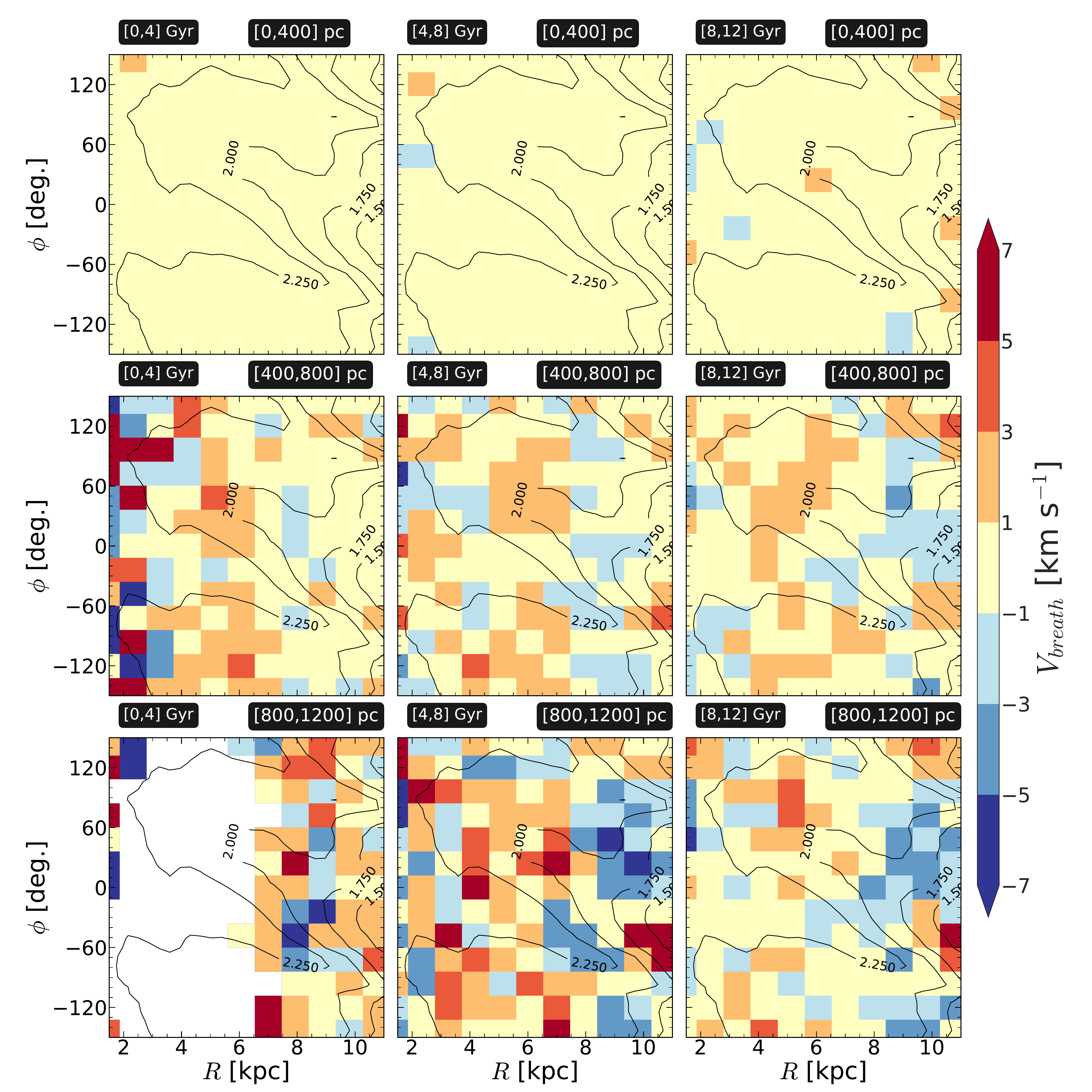}  
     
\caption{Star-forming model: Same as Fig.~\ref{fig:breath_xymap_nbodysph}, but shown in the $(R,\phi)$~plane. The age bins and the vertical layers are presented above each panel. Contours of {\it total} surface density (in log scale) are overlaid in each panel.  Large $|\vb|$ is associated with density peaks and troughs of the spiral.}
\label{fig:breath_Rphimap_nbodysph}
\end{figure*}

We therefore compute the breathing velocities for stars of different ages in three vertical slices: $|z| = [0, 400]\pc$, $|z| = [400, 800]\pc$, and $|z| = [800, 1200]\pc$. We calculate \vb\ using Eqn.~\ref{eq:vbreath} separately for each age population. Fig.~\ref{fig:breath_xymap_nbodysph} shows \vb\ in the $(x,y)$~plane whereas Fig.~\ref{fig:breath_Rphimap_nbodysph} shows \vb\ in the $(R,\phi)$~plane. Strong breathing motions are present in all three age populations; \vb\ has small amplitude near the mid-plane and increases with distance from it. This trend is present for all ages; at the largest heights the intermediate-age population has the strongest \vb\ amplitude, but the youngest population does not reach these heights so can only be compared in the intermediate layer. The intermediate population at the largest height shows that the compressive motions are closely associated to the peak density of the spiral in the outer region (excluding the central weak bar).

In summary, we find large-scale, non-zero vertical motions present both in the meridional plane as well as in the $(\phi,z)$~plane at a fixed radial annulus. These non-zero bulk vertical motions appear preferentially at azimuthal locations which contain the strongest spiral structure. The resulting breathing velocity, \vb, is small near the disc mid-plane, and increases with height, as found also by \citet{Debattista2014}. The breathing velocity in our model is largest for the intermediate age population at the largest heights, because the young population does not reach such heights; in the middle layer the young population has the largest \vb, while the old population has the smallest.

\section {A novel technique to detect the breathing motions}
\label{sec:novel_technique}
%&&&&&&&&&&&&&&&&&&&&&&&&&&&&&&&&&&&&&&&&&&&&&&&&&&&&&&

In the previous section, we measured the distribution of breathing velocity in the $(x, y)$ plane using Eq.~\ref{eq:vbreath}. This requires calculating, at a particular spatial location, the mean vertical velocity, $\avg {v_z}$, of stars in both the upper and the lower layers (of width $\Delta z$) of the disc, and then taking the difference of these two mean vertical velocities. For the simulated galaxy model we have the whole spatial coverage so that we can calculate quite robustly the spatial distribution of the resulting breathing velocities for stars at different vertical layers. However, for the Milky Way, this may not be the case due to the limitations in the spatial coverage and incompleteness of \gaia\ DR2.

Here, we develop a new technique for identifying breathing motions from the distribution of the mean vertical velocities (\avg {v_z}) as a function of height from the disc mid-plane ($z =0$). First, we demonstrate this technique on our star-forming model, and then apply it to the \gaia\ DR2 data. 

Above we showed that the breathing velocities increase away from the mid-plane.  Therefore to leading order, the variation of $\avg {v_z}$ with height can be fitted with a straight line. Breathing motions result in best-fit straight lines to the height variation of the mean vertical velocity that has a significantly \textit{non-zero} slope. If the slope is positive, then it indicates an expanding (positive) breathing motion, while if the slope is negative then it indicates the presence of a compressing (negative) breathing motion. 

\begin{figure*}
    \includegraphics[width=\linewidth]{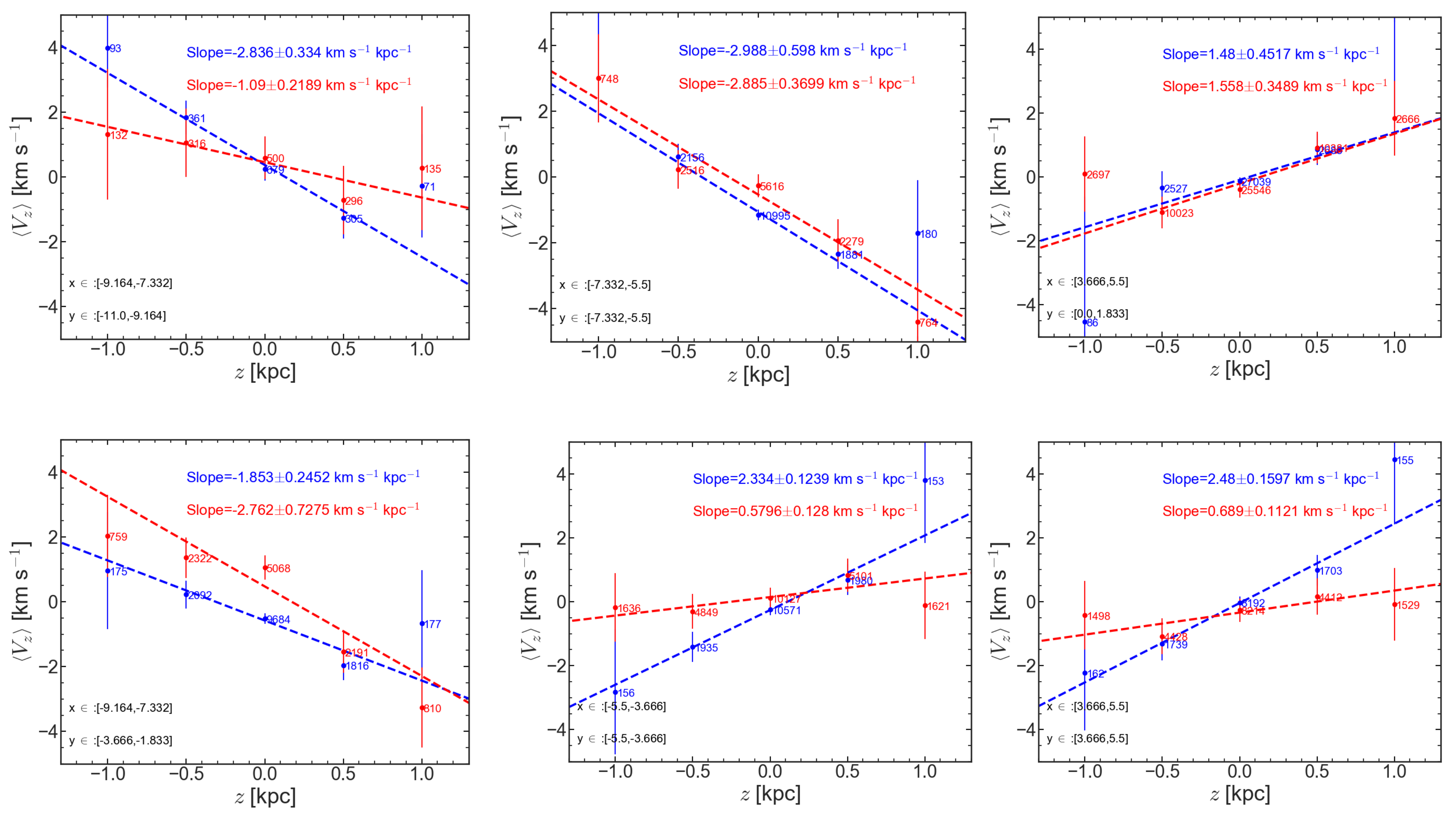}\par
\caption{ Star-forming model: a few examples of fitting straight lines to the variation of mean vertical velocity ($\left < v_z \right>$)  with height ($z$) for the younger (shown in blue) and the older population (shown in red) are shown. The resulting best-fit slopes and the associated errors are indicated in each sub-panel. The values listed near each point denote the total number of stellar particles present in that bin. The (absolute) values of the best-fit slope for the younger population are, in general, larger than that for the older population.}
\label{fig:strlinefit_sim}
\end{figure*}

To test this, we divide the stellar particles into two age populations: younger ($1-6 \Gyr$), and older ($6-12 \Gyr$) stellar particles. We exclude the very young stellar particles (age less than $1 \Gyr$) because of their limited vertical coverage. Fig.~\ref{fig:strlinefit_sim} shows a few examples of fitting a straight-line to the variation of $\avg {v_z}$ with height. The straight-line is fitted via the {\sc scipy} package {\sc curvefit} which uses the  Levenberg-Marquardt algorithm.
The non-zero slope indicates breathing motions with amplitude increasing with height from the mid-plane. The (absolute) value of the slope is larger for the younger stellar population than for the older stellar population. This implies that the breathing motions are stronger in the younger population than the older populations, in agreement with the findings of the previous section. We note that, at certain locations, there is a clear difference in the zero-point of the different age groups. When studied in detail, we found that these zero-point offsets arise because of weak bending waves, which have different amplitudes in the different age groups (Khachaturyants et al. {\it submitted}). 

\begin{figure}
    \includegraphics[width=1.025\linewidth]{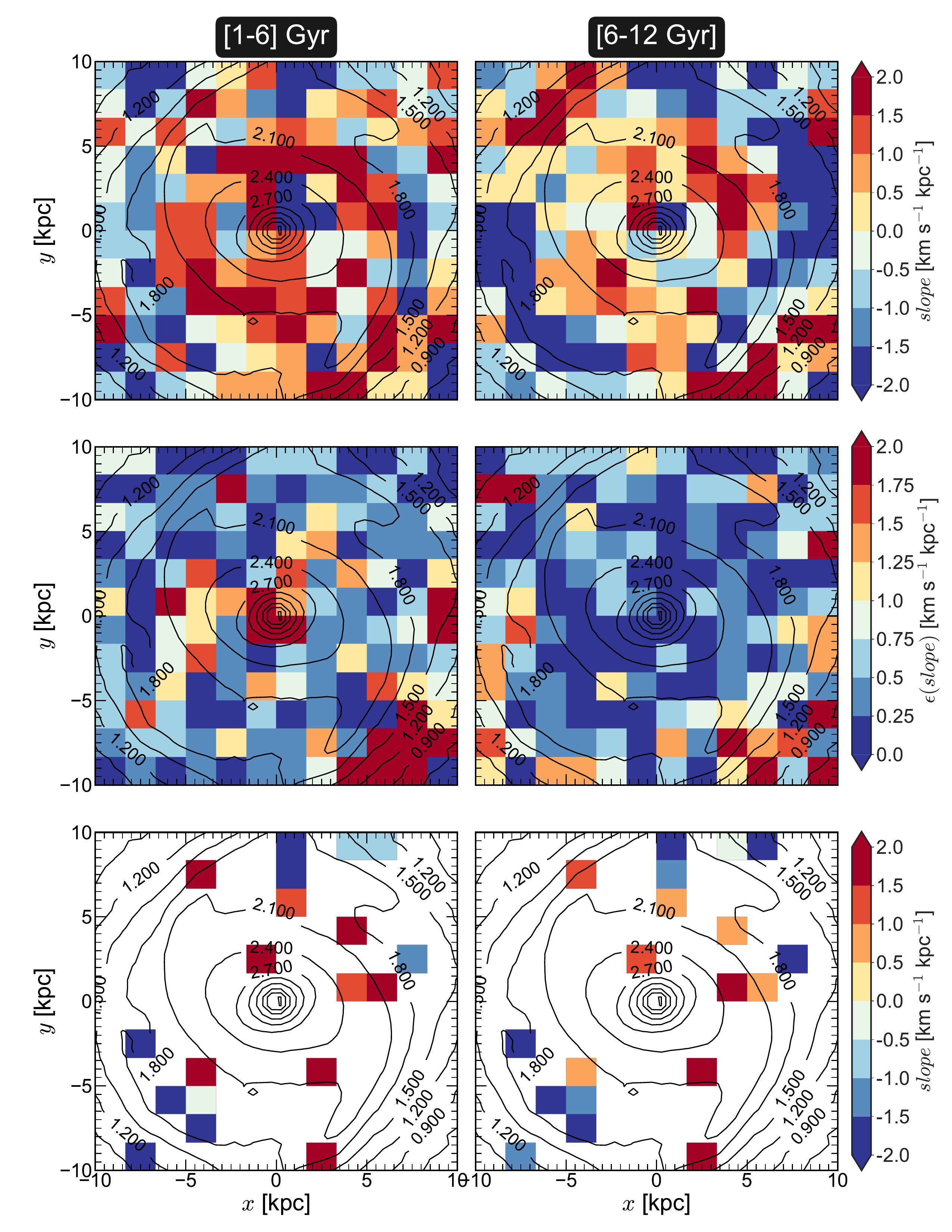}\par
\caption{ Star-forming model: distributions of the best-fit slope (top panels) and the associated errors (middle panels) in the $(x, y)$~plane for two populations with different ages are shown. Bottom panels show the same distribution of the best-fit slope, but only for those bins where a simultaneous quality cut of $slope/\epsilon (slope) > 3$, for both the stellar populations, is met. Black solid lines denote the contours of constant density.}
\label{fig:strlinefit_simXYmaps}
\end{figure}

We compute the error on the slope ($\epsilon(slope)$) and plot this in the middle row of Fig.~\ref{fig:strlinefit_simXYmaps}. For uniform comparison, we impose a simultaneous quality cut of $slope/\epsilon (slope) > 3$ for both age populations, and choose only those bins where this quality-cut is met. The resulting distribution of the best-fit slopes is shown in the bottom panels of Fig.~\ref{fig:strlinefit_simXYmaps}. Clear signatures of breathing motions (with amplitude increasing with height) are seen. The best-fit slopes for the younger population ($1-6 \Gyr$) are, in general, larger than those for the older population ($6-12 \Gyr$).

\begin{figure}
    \includegraphics[width=1.025\linewidth]{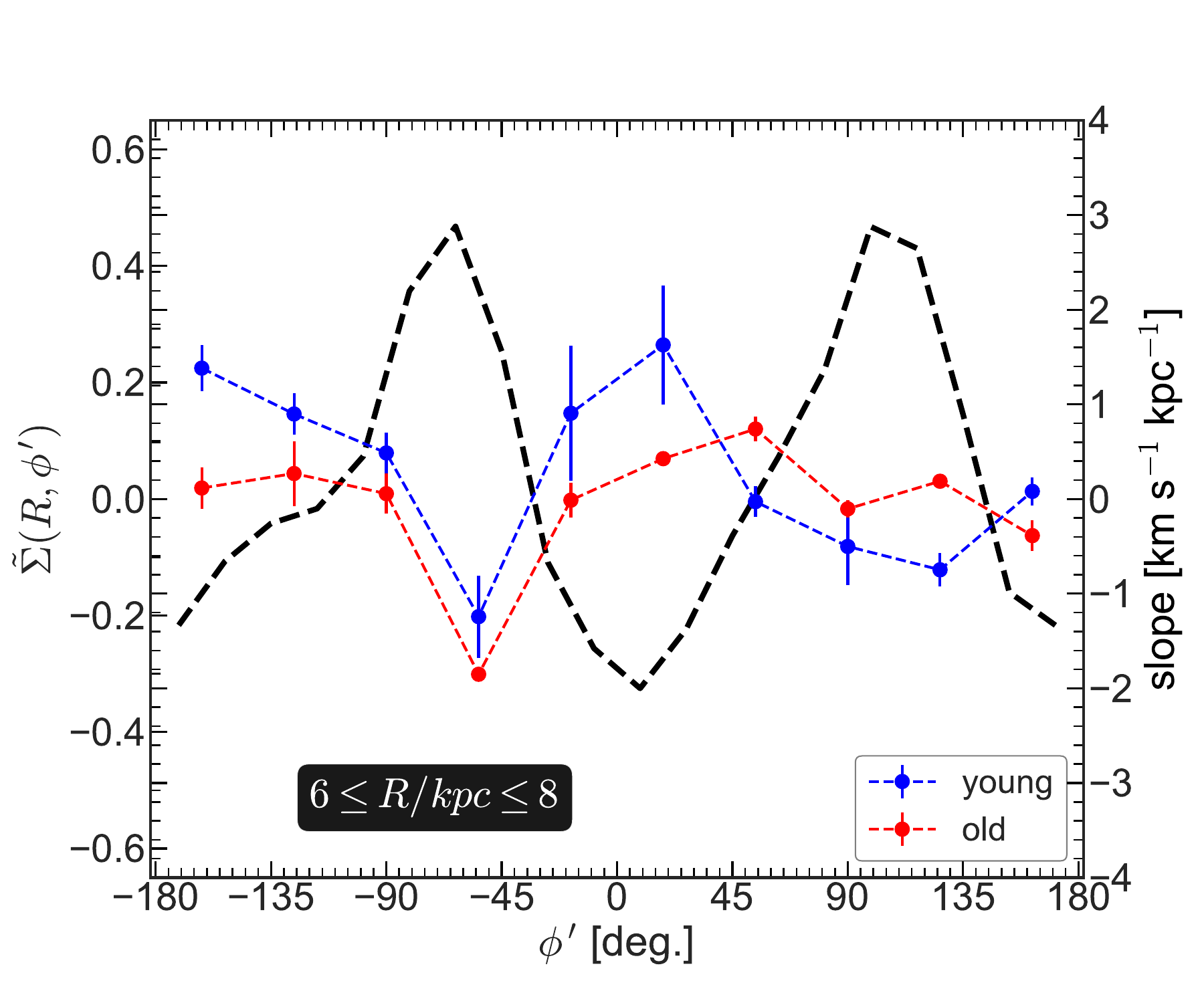}\par
\caption{Star-forming model: variation of the slopes for the young (blue circles) and the old stellar populations (red circles) are shown as a function of the rotated azimuthal angle ($\phi'$). Only the stellar particles in the radial range $6 \kpc$ to $8 \kpc$ are chosen here. The black dashed line denotes the azimuthal variation of the total residual surface density ($\tilde \Sigma (R, \phi')$), calculated in this chosen radial extent. The particles have first been binned in $1~\kpc$ annuli and then azimuthally rotated relative to each other so that the minimum in $\tilde \Sigma (R, \phi)$) in each annulus is coincident. Only then are the vertical slopes computed, amounting to stacking different radial ranges while unwinding the spiral. The sense of rotation is towards increasing $\phi'$.}
\label{fig:compare_slopevsazimuth_sim}
\end{figure}

Lastly, we investigate how the slope varies as a function of the azimuthal angle for the young and the old populations relative to the density variation. To achieve this, we first choose a radial extent ranging from $6 \kpc$ to $8 \kpc$ (\ie\ outside the CR of the spiral), where the model exhibits a strong spiral. It is evident from Fig.~\ref{fig:density_residualmaps_nbodysph} (bottom panels) that the azimuthal locations of the density peaks vary as a function of radius in the chosen radial extent, because of the spiral's pitch angle. Therefore, to obtain a stronger signal of the slope, we rotate the stars in different $1\kpc$-width radial bins, in such a way that the density peaks in our chosen radial extent coincide. Then, we recalculate the slope of the breathing velocity as a function of the rotated azimuthal angle ($\phi'$) for the young and the old stellar populations. This is shown in Fig.~\ref{fig:compare_slopevsazimuth_sim}. We also calculate the total residual surface density ($\tilde \Sigma(R, \phi')$) as a function of the rotated azimuthal angle in this chosen radial extent. As evident from Fig.~\ref{fig:compare_slopevsazimuth_sim}, the locations of the larger slopes of the breathing motions coincide with the azimuthal locations of the density minima while the azimuthal locations  of the maximum density exhibit smaller values of the slope. In other words, the amplitudes of the breathing motions are larger in the inter-arm region. This trend is consistent with the earlier findings of \citet{Debattista2014} that stars on the expanding side of the spiral always exhibit larger breathing motions when compared with those for the compressing side of the spiral. \citet{Debattista2014} attributed this behaviour to the more abrupt density variation as stars leave the spirals compared with when they enter them. (The test particle and semi-analytic calculations of \citet{Faureetal2014} did not exhibit this behaviour because the spiral perturbation they used had a purely $m=2$ multiplicity.) Furthermore, the slope of the young stellar population is larger than the slope for the older stellar population, thereby demonstrating that indeed the breathing motion is stronger in the young stellar population than in the old stellar population.

\section{Comparison with vertical motions in the Milky Way}
\label{sec:gaia_comparison}
%&&&&&&&&&&&&&&&&&&&&&&&&&&&&&&&&&&&&&&&&&&&&&&&&&&&&&&

We now search for these signatures of spiral-driven breathing motions in the \gaia\ DR2 data as a function of stellar ages, and compare with the results obtained from the star-forming model.

\subsection{Sample selection from the Sanders \& Das dataset}
\label{sec:gaia_sample_selection}
%&&&&&&&&&&&&&&&&&&&&&&&&&&&&&&&&&&&&&&&&&&&&&&&&&&&&&&&&

We use the publicly available sample of stellar ages of \citet{SandersandDas2018} (hereafter, SD18), which provides a catalogue of distance, mass, and age for approximately 3 million stars in the \gaia\ DR2. 
This sample allows us to study the vertical motions for different age populations. \gaia\ DR2 suffers from distance biases from the parallax measurements \citep[for detailed discussion, see e.g.,][]{Bailer_Jones_2018,Lurietal2018,Lindegrenetal2018,Schoenrichetal2019}. To account for these biases, especially when the parallax measurements are of the same order as the associated uncertainties, SD18 used a full Bayesian framework to calculate the distance of a star while using different photometric and spectroscopic parameters from a number of ground-based surveys (APOGEE, \gaia-ESO, GALAH, LAMOST, RAVE, and SEGUE), and the astrometric information from \gaia\ DR2, simultaneously as priors \citep[for details see section~2 of][]{SandersandDas2018}. Similarly, the probability distributions for mass, and ages for each star were calculated by SD18 via a Bayesian framework based on the photometric, spectroscopic, and astrometric quantities. This dataset includes the six-dimensional Galactocentric cylindrical position-velocity coordinates, with the Sun placed at a Galactocentric radius $R_0 = 8.2 \kpc$ and a height above the plane $z_0 = 15 \pc$ \citep{Bland-Hawthorn2016}, and the values of the Solar peculiar motions from \citet{schronrichetal2010}. 
\par
Each stellar entry in the SD18 dataset is characterised by three flags that, depending on the nature of the analysis, can aid in filtering out problematic data. The {\sc flag} entry characterises all issues related to the quality of the input data (spectroscopy, astrometry, photometry) from the above listed surveys and of the output data from the SD18 Bayesian frameworks (mass, age). If there are no issues in the input or output data, then {\sc flag} $=0$. The {\sc duplicated} flag specifies if stellar entries are duplicates due to a crossmatch within a single survey or between multiple surveys. Within a single survey, the duplicate with the smallest vertical velocity uncertainty is kept, while duplicates between surveys are kept based on the {\sc flag} entry and a survey hierarchy set in SD18 (APOGEE, GALAH, GES, RAVE-ON, RAVE, LAMOST and SEGUE). For example, if APOGEE and RAVE contain a duplicate entry with proper input and output data ({\sc flag} $=0$) then, regardless of the relative errors, the APOGEE entry is kept and labelled as {\sc duplicated} $=0$. Lastly, the {\sc best} flag in SD18 encompasses the two prior flags and is 1 when {\sc flag} $=0$, {\sc duplicated} $=0$ and the star has a valid cross-match in \gaia\ DR2.

For this work, we choose a sub-sample of stars from the SD18 dataset by applying a number of quality cuts. In particular, we choose only those stars for which distance ($d$) is less than $10 \kpc$, $1/parallax < 10 \kpc$, {\sc best} $=1$, and the ratio of parallax to error in parallax, $\pi/\sigma_\pi > 3$. Since the parallax/distance bias is more significant for very young stars \citep[for details see][]{SandersandDas2018}, we discard all the stars with ages less than $1 \Gyr$. Our selected sub-sample has a total of 3,147,069 stars.

\subsection{Comparison of SD18 dataset with a bias-controlled dataset}
\label{sec:com_SD18_SME19}
%&&&&&&&&&&&&&&&&&&&&&&&&&&&&&&&&&&&&&&&&&&&&&&&&&

\begin{figure*}
   \includegraphics[width=\linewidth]{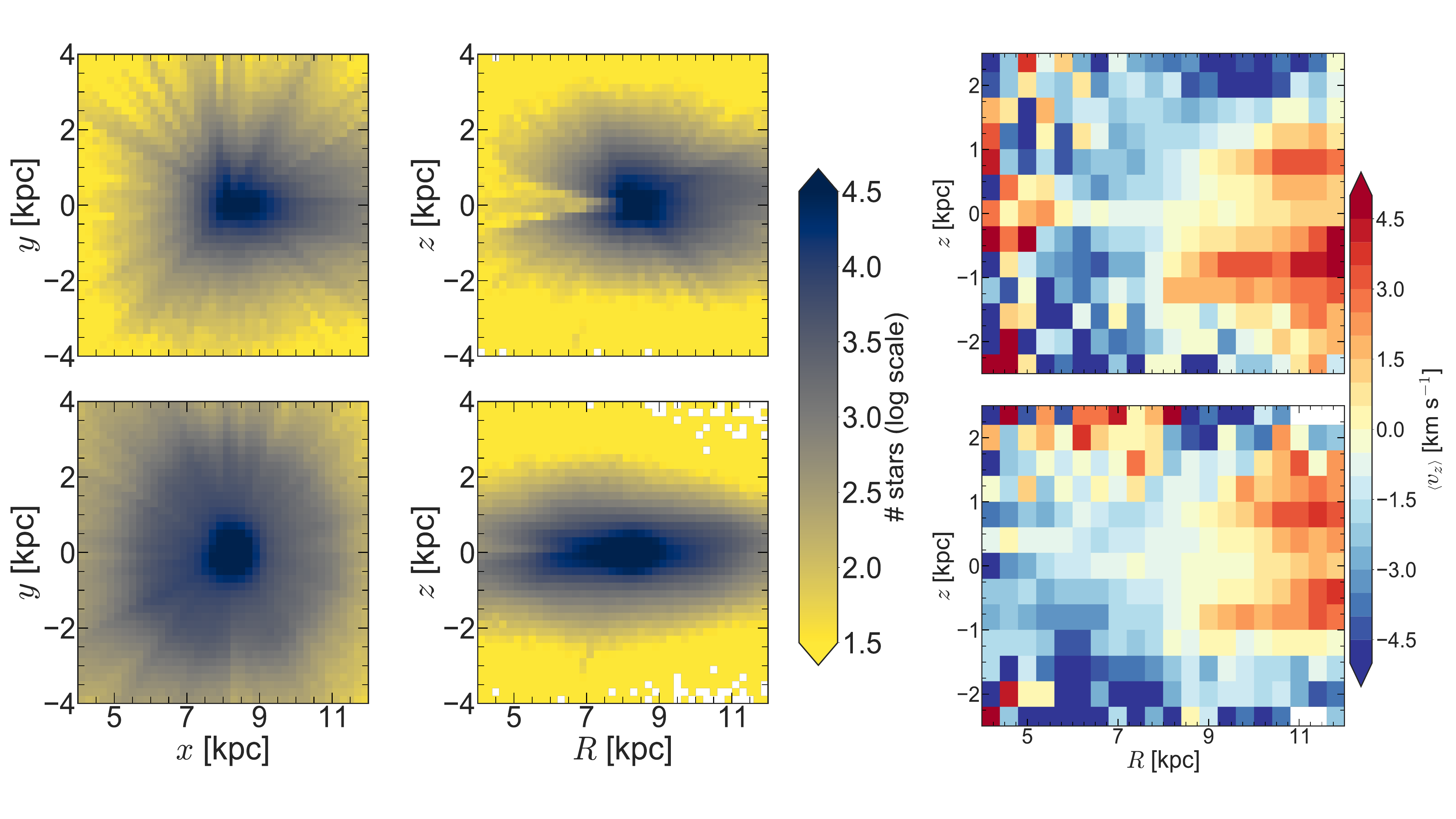}
\caption{Comparison between the SD18 and SME19 datasets: density distribution of stars in the $(x,y)$-plane (left column) and in the $(R,z)$-plane (middle column), and the mean vertical velocity in the $(R,z)$-plane (right column) are shown for the SD18 (top panels) and the SME19 (bottom panels) datasets. For the details of the quality cut applied to both datasets, see section~\ref{sec:com_SD18_SME19}.}
\label{fig:compare_sd18_sme18}
\end{figure*}

Before we proceed to the calculation of breathing motions for the Milky Way using our selected sub-sample from the SD18 dataset, first we compare our sub-sample with a bias-controlled dataset for the Milky Way. We choose the  \citet{Schoenrichetal2019} dataset (hereafter, SME19) who corrected for the bias in the \gaia\ parallax, and derived the distances of all stars in the RVS (Radial Velocity Spectrograph) sample of the \gaia \ DR2 via a Bayesian framework \citep[for details, see][]{Schoenrichetal2019}. 

\begin{figure}
   \includegraphics[width=1.02\linewidth]{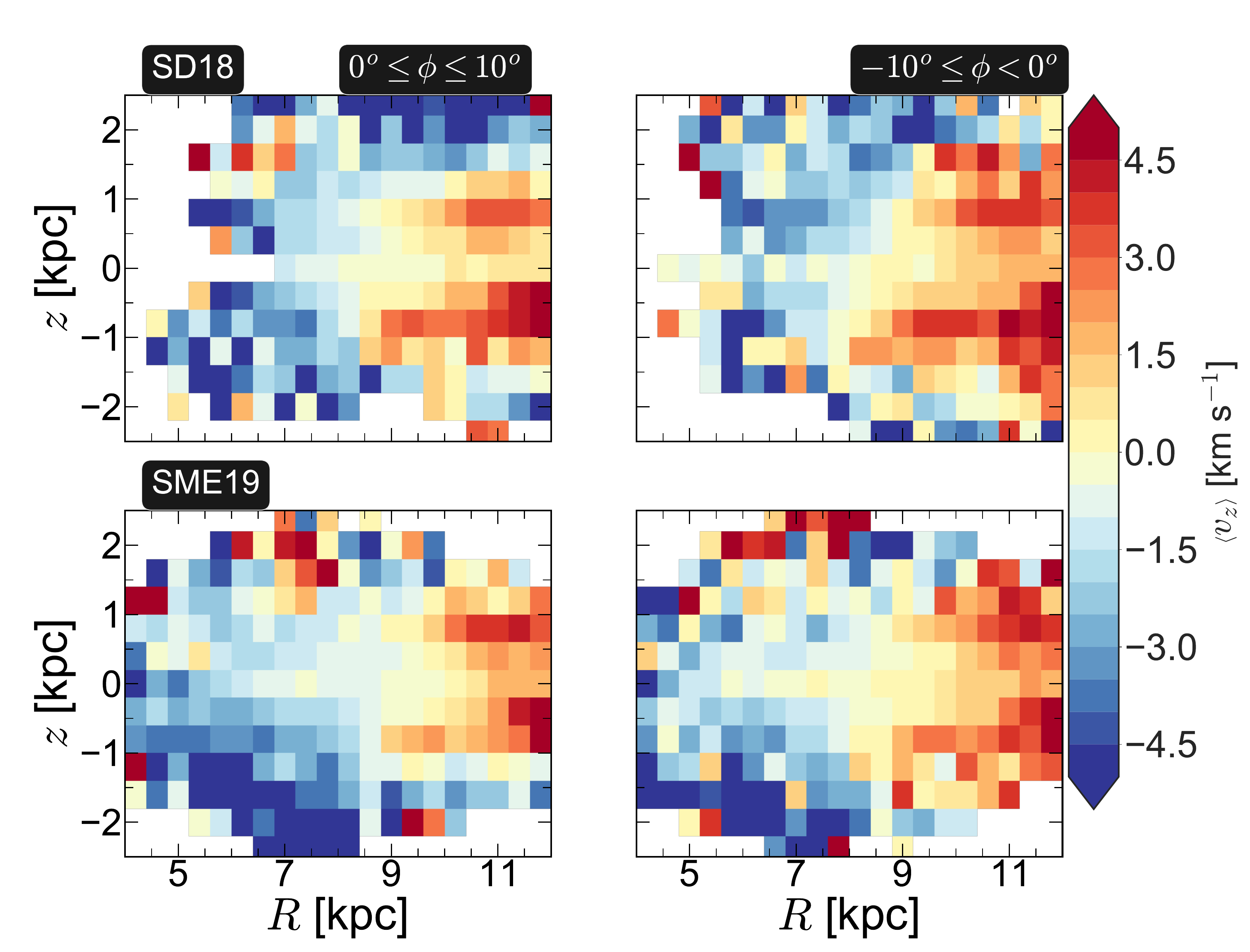}
\caption{Comparison between the SD18 and SME19 datasets: distributions of the mean vertical velocity, $\avg{v_z}$, are shown for our selected samples of stars when they are split into positive and negative azimuthal angles. The top panels show results for the SD18 dataset while the bottom panels show those for the SME19 dataset. The extent of azimuthal range is provided at the top of each column.}
\label{fig:compare_meanvz_sd18_sme18_phicut}
\end{figure}

\begin{figure}
   \includegraphics[width=1.02\linewidth]{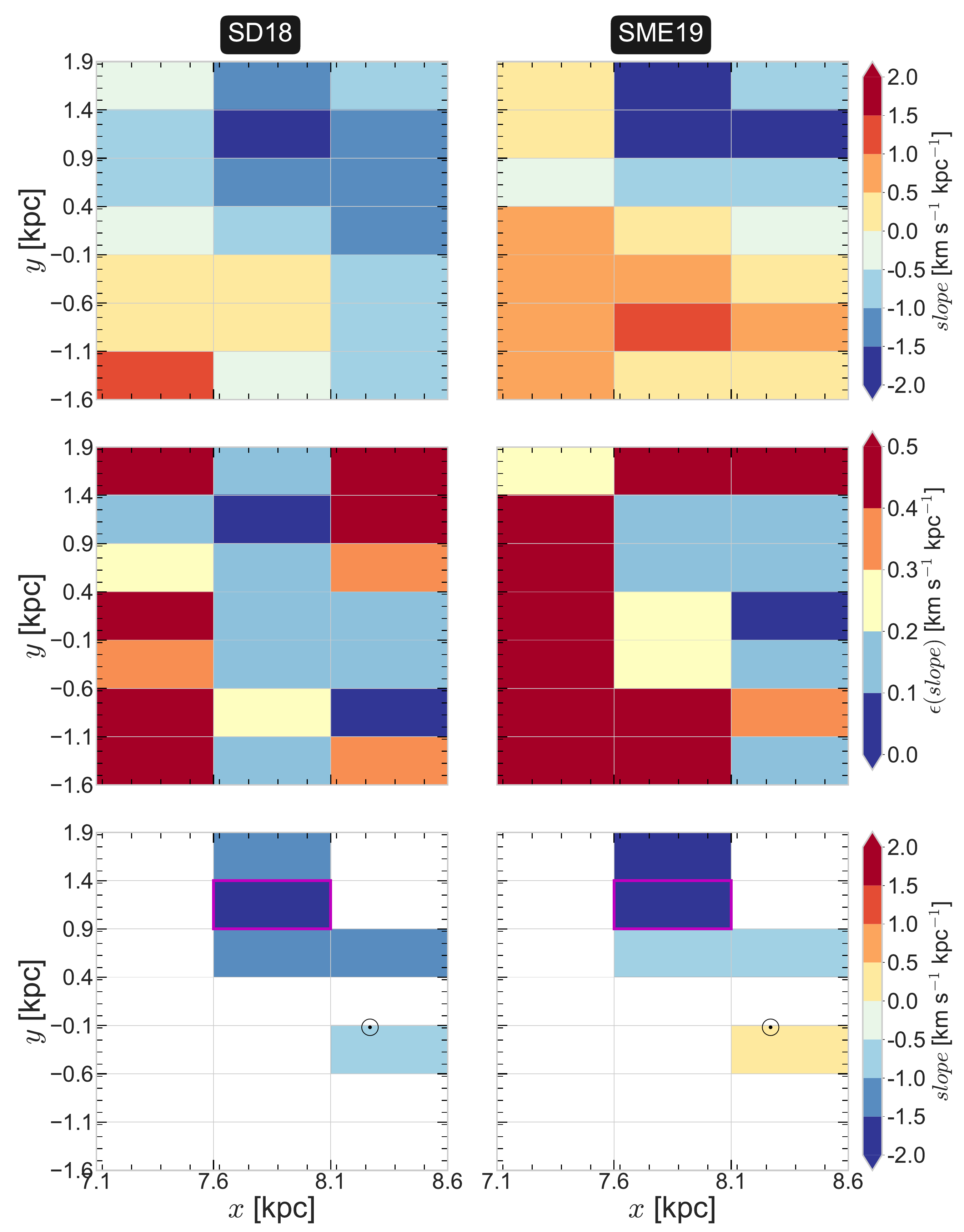}
\caption{Comparison of the breathing motions in the SD18 and SME19 datasets : distributions of the best-fit slope (top panels) and the associated errors (middle panels) in the $(x, y)$~plane are shown for the SD18 (left panels) and the SME19 (right panels) datasets. Bottom panels show the same distribution of the best-fit slope, but only for those bins where a simultaneous quality cut of $slope/\epsilon (slope) > 3$, for both the datasets, is met separately. The magenta rectangles in the bottom panels denote the chosen spatial bin where we study the age variation of the breathing motions (see text for details). The Sun's position is also indicated in the bottom panels.}
\label{fig:compare_Vbreath_sd18_sme18_xydistribution}
\end{figure}

\begin{figure}
   \includegraphics[width=1.02\linewidth]{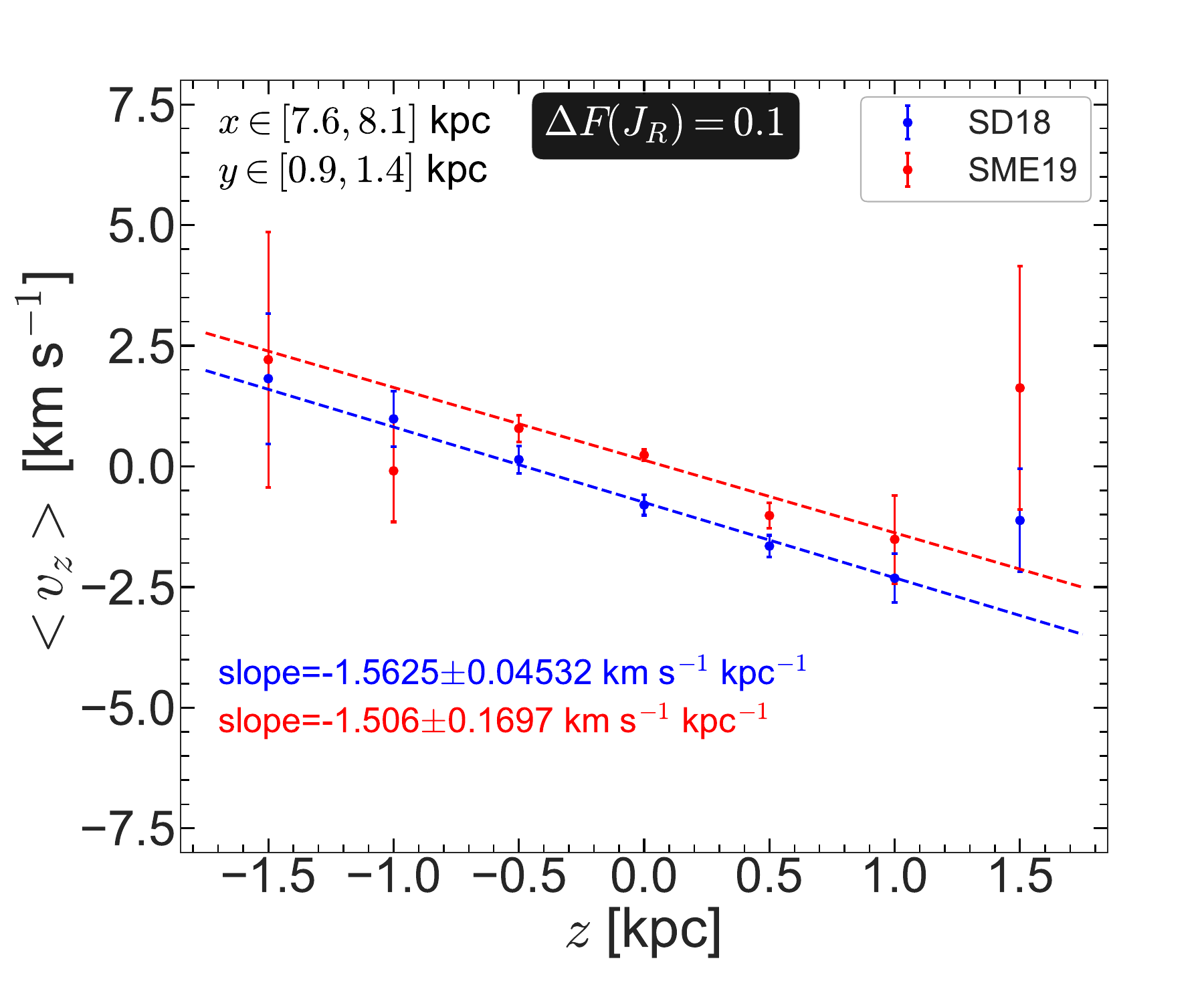}
\caption{Comparison of the breathing motions in the SD18 and SME19 datasets. The best-fit straight line fits to the $\avg{v_z}$ versus $z$ data are shown for both datasets. The spatial location where the slopes are measured, indicated at the top of the figure, is chosen to maximise the slope in the SD18 dataset. The corresponding best-fit slope and the associated error are indicated in the top left label. Details of the quality cuts applied to both datasets are given in section~\ref{sec:com_SD18_SME19}. The error bars on $\avg{v_z}$ are the standard errors ($\sigma_{v_z}/\sqrt{N}$) on the mean vertical velocity.}
\label{fig:compare_Vbreath_sd18_sme18}
\end{figure}

Since the SME19 dataset does not include stellar ages, we do not include any age cuts on the SD18 dataset for this comparison. In both datasets we select stars with distance $ <10 \kpc$ and with the ratio of parallax to parallax error, $\pi/\sigma_\pi > 3$.
Furthermore, to reduce the contamination from both very young stars and halo stars (at larger heights), we further constrain the sample based on the radial action ($J_R$) of the individual stars. The SD18 dataset includes the radial action of individual stars; for the SME19 dataset we calculate these ourselves. To calculate $J_R$, we use the St{\"a}ckel fudge method \citep{Binney12,Sanders+16} from the {\sc AGAMA} software library \citep{agama} in the potential of \cite{McMillan}. The St{\"a}ckel fudge function in {\sc AGAMA} takes the Galactocentric 6D coordinates of the SME19 dataset and produces the full action-angle coordinates, including the $J_R$ values.

We then compute the cumulative distribution function (CDF) of the radial action, $F(J_R)$, of the SME19 dataset. The very young stars are, in general, on almost circular orbits and thus should have small values of $J_R$. On the other hand, the halo stars at larger heights should have large values of $J_R$. Therefore, we choose stars from the SME19 dataset, while rejecting the top and bottom 10 percent  of stars in $F(J_R)$. We denote this rejection process as $\Delta F(J_R) =0.1$. To make a uniform comparison, we apply the same radial action cut ($\Delta F(J_R) =0.1$) on the SD18 dataset using the same values of upper and lower $J_R$ as in the SME19 dataset. Fig.~\ref{fig:compare_sd18_sme18} shows the corresponding distributions of stars in the $(x, y)$ plane and in the $(R, z)$ plane, as well the distribution of mean vertical velocity, $\avg {v_z}$, in the $(R, z)$ plane for both the SD18 and SME19 datasets. We split our selected samples of stars from both the datasets, into positive and negative Galactocentric azimuthal angles, and recalculate the distribution of $\avg{v_z}$, which is shown in Fig.~\ref{fig:compare_meanvz_sd18_sme18_phicut}. The distribution of the mean vertical velocity ($\avg {v_z}$) in the $(R, z)$ plane, calculated for both the SD18 and SME19 datasets appear to match quite well. Further, we checked that a more conservative cut on $J_R$, e.g. $\Delta F(J_R) =0.15$, does not alter the main findings presented here and in the subsequent sections.

We calculate the distributions of the bending velocity ($\vbend$) in three vertical layers: $|z| = [0, 0.4] \kpc$, $|z| = [0.4, 0.8] \kpc$ and $|z| = [0.8, 1.2] \kpc$ (using Eq.~\ref{eq:vbend}) for both the SD18 and SME19 datasets. The resulting bending velocities calculated from the two datasets are broadly in agreement;  these results are not shown here but are evident in Fig.~\ref{fig:compare_sd18_sme18}. 
We also compare the breathing motions, $\vb$, between the two datasets. Because the breathing motions are significantly smaller than the bending motions, we start looking for the signature of the breathing motions in the Solar Neighbourhood in the following way. First, we calculate the distribution of the best-fit slope in the $(x,y)$ plane around the Solar Neighbourhood, by fitting a straight line to the variation of $\avg {v_z}$ with $z$, in a similar way as done for the model. The corresponding distribution of the best-fit slope, and the associated error ($\epsilon(\rm slope)$) in the $(x,y)$ plane, for both the SD18 and the SME19 datasets are shown in Fig.~\ref{fig:compare_Vbreath_sd18_sme18_xydistribution} (see top and middle panels). The bottom panels of Fig.~\ref{fig:compare_Vbreath_sd18_sme18_xydistribution} show the bins for which a quality-cut of $slope/\epsilon(slope) >3$ is met \textit{simultaneously} for these datasets. We find that the location where we get the most consistent signature of the  \vb\ is $x \in [7.6, 8.1] \kpc$ and $y \in [0.9, 1.4] \kpc$. In this bin, the best-fit slope values, obtained from the two datasets agree with each other (within their error-bars). Moreover, we find that this bin shows the largest value for the best-fit slope, \ie\ the signature of the largest \vb. The presence of this `consistent' signature of \vb\ is not seen for any other bins shown in the bottom panels of Fig.~\ref{fig:compare_Vbreath_sd18_sme18_xydistribution}. Hence, we will only consider this spatial bin for the subsequent study of the age-dependence of the \vb\ in the Solar Neighbourhood. The corresponding slope of \avg{v_z} versus $z$ at the same spatial location using the two datasets are shown in Fig.~\ref{fig:compare_Vbreath_sd18_sme18}. We note that there is an offset between the $\avg{v_z}$ values calculated from both the datasets, especially near the disc mid-plane; this could possibly be due to the different zero-points used in the datasets. The values of the slopes obtained from the two datasets match quite well within their error-bars. In both cases, the $slope/\epsilon(slope)$ is found to be greater than 3 so that in both datasets the detection of \vb\ is significant.

\subsection{Vertical motions for different ages in the Milky Way}
\label{sec:gaiaVertMotion_agecut}
%&&&&&&&&&&&&&&&&&&&&&&&&&&&&&&&&&&&&&&&&&&&&&&&&&&&&&&&

We now study the non-zero mean vertical motion, and the associated breathing motions, as a function of the stellar ages in the same volume as where we found the largest breathing motions. We can do this only using the SD18 dataset.

\begin{figure}
%\begin{multicols}{2}
    \includegraphics[width=\linewidth]{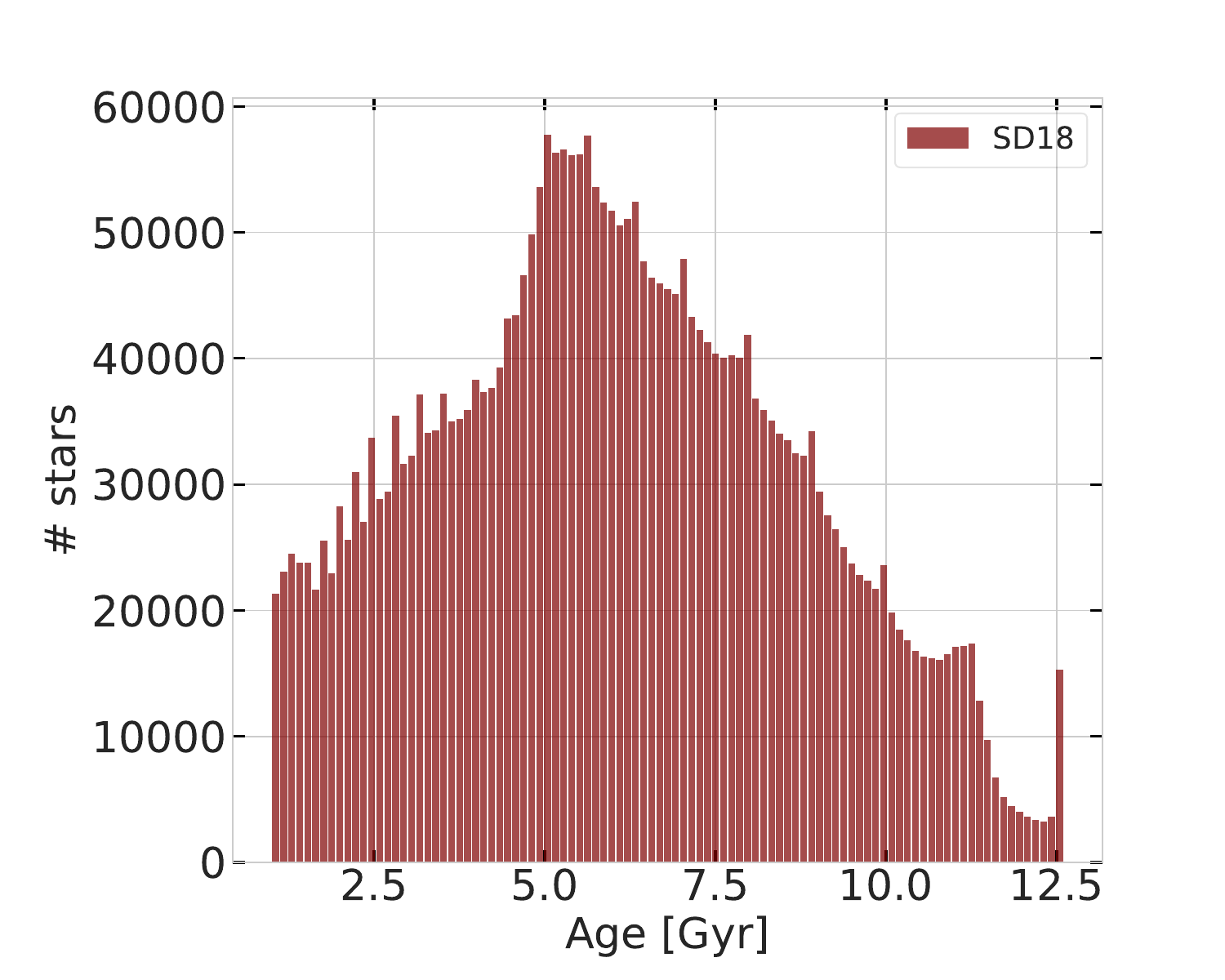}
    \medskip
    \includegraphics[width=\linewidth]{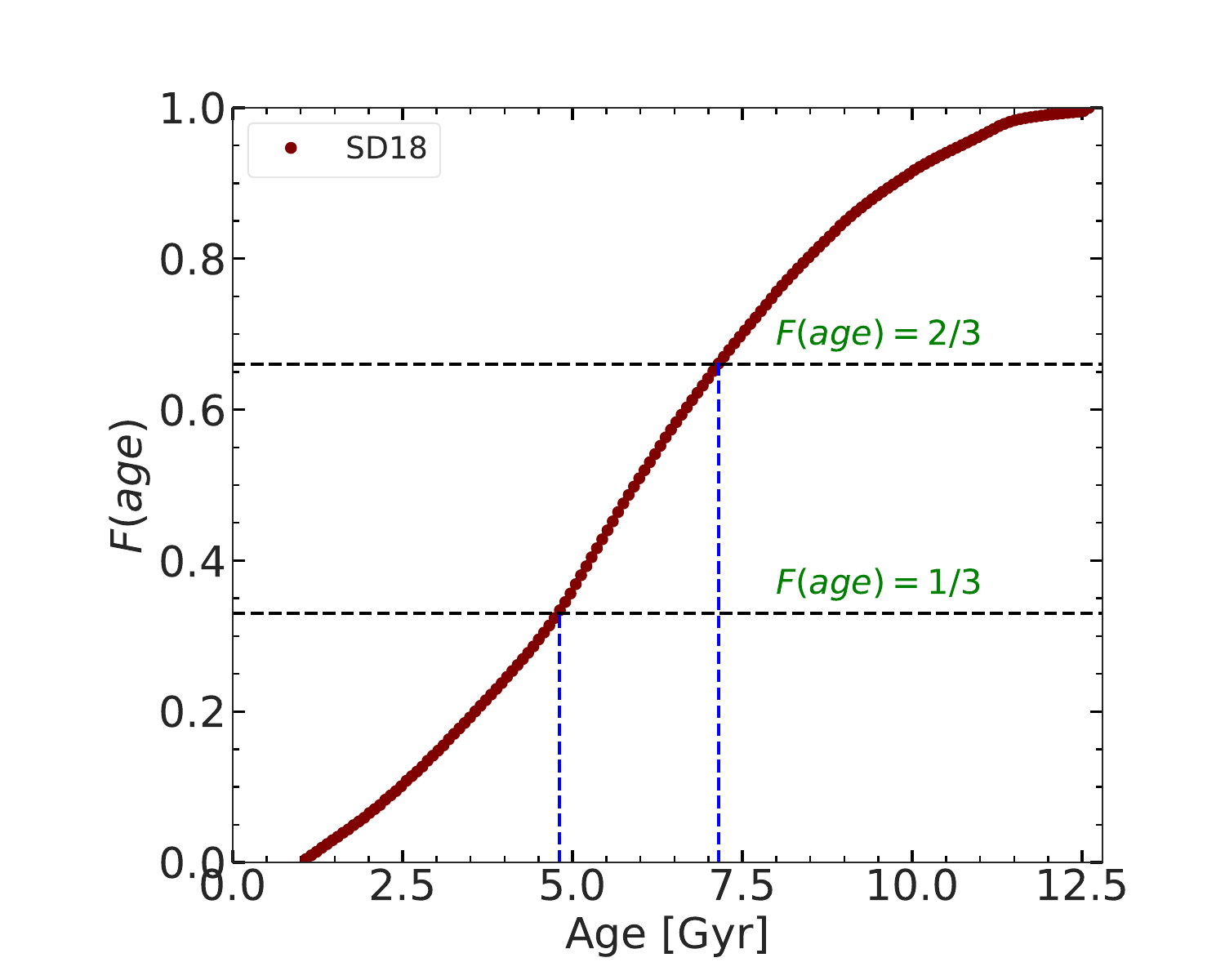}
%    \end{multicols}
\caption{ SD18 sample: histogram of stellar ages ({\it top panel}) and the corresponding cumulative distribution function (CDF), $F(\rm age)$ of stellar ages ({\it bottom panel}) of our selected sample from the SD18 dataset. The black dashed lines in the {\it bottom panel} indicate the cuts applied on the CDF whereas the vertical blue dashed lines indicate the corresponding stellar ages.}
\label{fig:appen_sd18_agedistrinution}
\end{figure}

\begin{figure*}
    \includegraphics[width=1.05\linewidth]{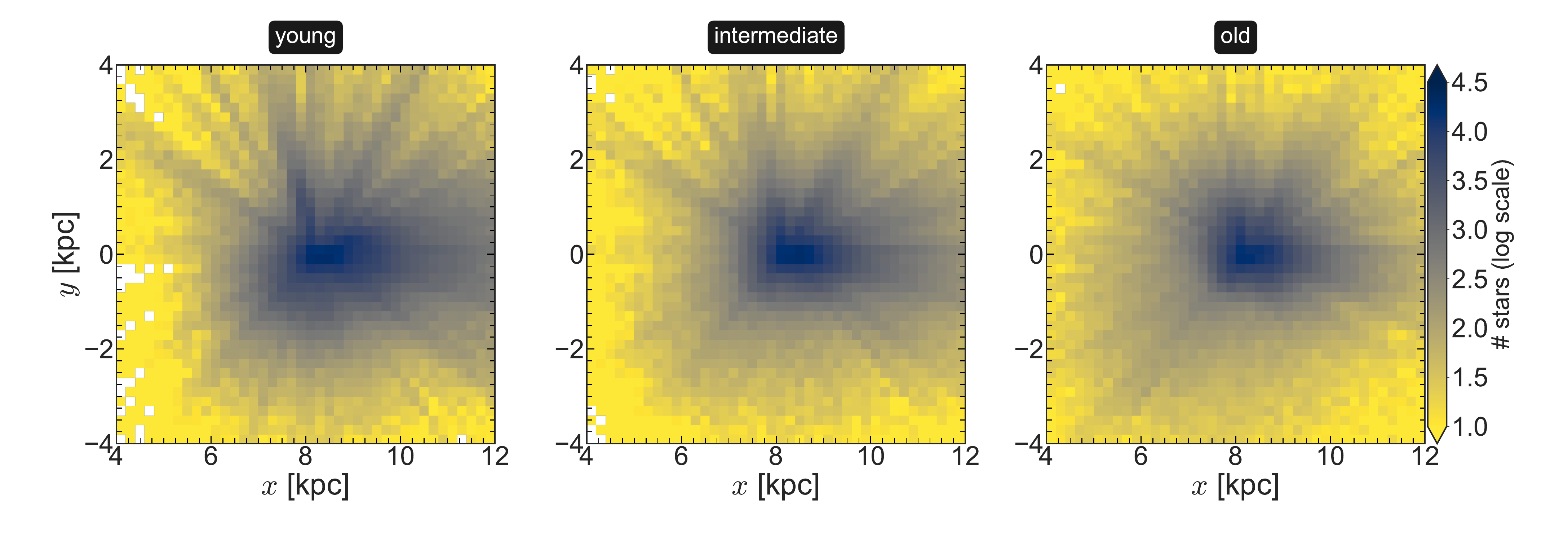}
    \medskip
        \includegraphics[width=1.05\linewidth]{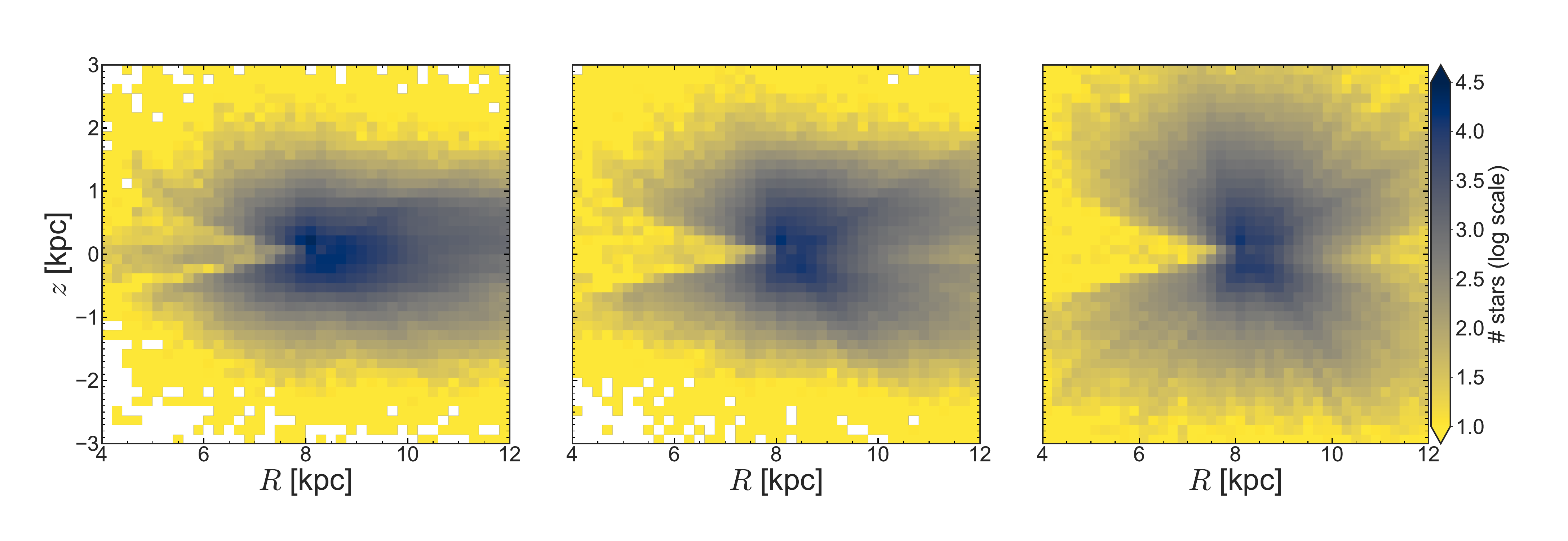}
\caption{SD18 sample: distribution of our selected sample from SD18 is shown in the $(x, y)$~plane ({\it top panels}) and in the $(R, z)$~plane ({\it bottom panels}) for three stellar populations: young ({\it left panel}), intermediate ({\it middle panel}), and old ({\it right panel}) stellar populations (for details see section~\ref{sec:gaiaVertMotion_agecut}). An additional cut on the radial action ($J_R$) is applied; for details see text in section~\ref{sec:com_SD18_SME19}.}
\label{fig_appen:age_cut_gaiasd}
\end{figure*}

Fig.~\ref{fig:appen_sd18_agedistrinution} presents how the selected sample of stars from the SD18 dataset is divided into three equal-sized populations by cutting the cumulative distribution function, $F$(age), of the stellar ages. The resulting sub-populations are: young ($0 \le F(\rm age) < 1/3$), intermediate ($1/3 \le F(\rm age) < 2/3$), and old ($2/3 \le F(\rm age) < 1$) stellar populations, or, in terms of the stellar ages, $[1, 4.85]~\Gyr$ (young), $[4.85, 7.1]~\Gyr$ (intermediate), and $[7.1, 12.6]~\Gyr$ (old). Fig.~\ref{fig_appen:age_cut_gaiasd} shows  the distributions of these stellar populations in the Galactocentric $(x, y)$~plane, as well as in the $(R, z)$~plane.

\begin{figure*}
       \includegraphics[width=\linewidth]{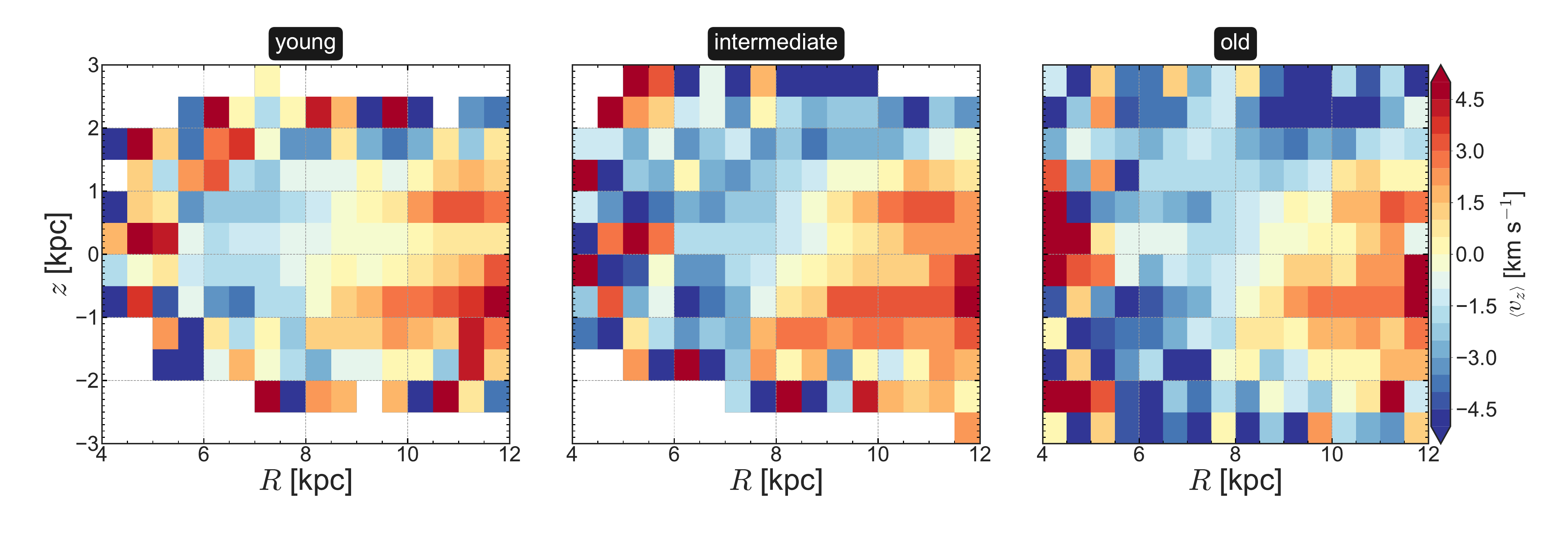}
 \caption{SD18 sample: mean vertical velocity, $\avg{v_z}$, for different stellar populations in the $(R, z)$~plane. From {\it left to right}: stars with ages $1-4.85~\Gyr$ (young), $4.85-7.1~\Gyr$ (intermediate), and $7.1-12.6~\Gyr$ (old), are shown (for details see section~\ref{sec:gaiaVertMotion_agecut}). An additional cut on the radial action ($J_R$) is applied; for details see text in section~\ref{sec:com_SD18_SME19}.}
 \label{fig:gaia_bulkmotions_agecut}
 \end{figure*}

Fig.~\ref{fig:gaia_bulkmotions_agecut} shows the mean vertical velocity distribution in the $(R,z)$~plane for the three stellar populations. Non-zero mean vertical motion is seen in all three stellar populations. Stars on either side of the mid-plane are seen to move in opposite directions, \ie\ with a non-zero \vb\ at larger heights, while at moderate heights all populations exhibit bending motions (\ie\ stars on either side of the mid-plane are moving in the same direction). 

\begin{figure}
       \includegraphics[width=1.02\linewidth]{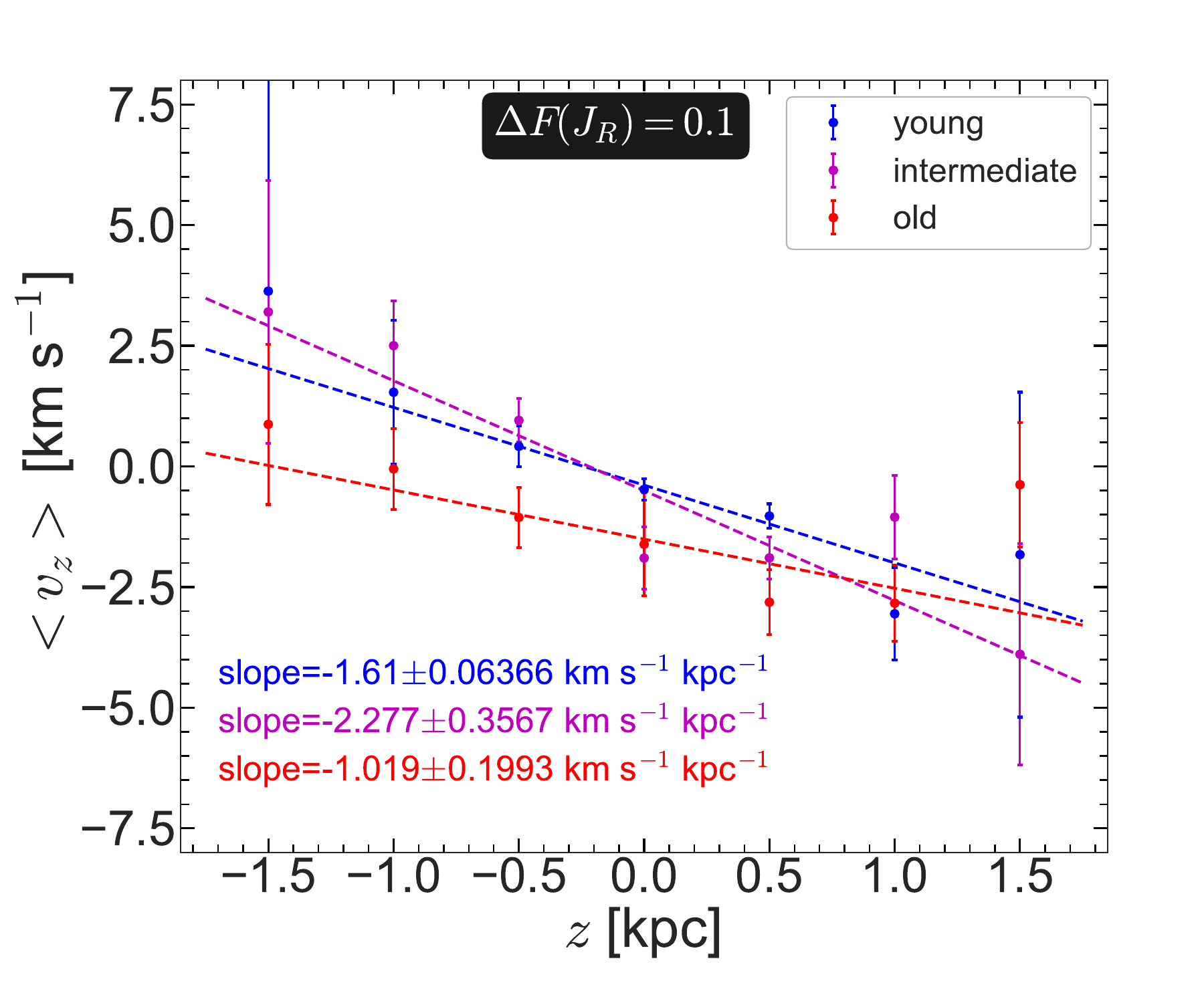}\par
 \caption{SD18 sample: The best-fit straight lines to the $\left < v_z \right>$ versus $z$ data are shown for three populations with different ages, calculated at the spatial location $x \in [7.6, 8.1] \kpc$ and $y \in [0.9, 1.4] \kpc$. The corresponding best-fit slope and the associated error are mentioned in the figure. The intermediate stellar population has the largest slope while the old stellar population shows the smallest slope. The error bars on $\avg{v_z}$ are the standard errors ($\sigma_{v_z}/\sqrt{N}$) on the mean vertical velocity.}
 \label{fig:gaia_strlinefit}
 \end{figure}

To measure the variation of the breathing velocity as a function of stellar age, we consider the same spatial location as before, i.e. $x \in [7.6, 8.1] \kpc$ and $y \in [0.9, 1.4] \kpc$, where a strong (and consistent) breathing motion is present without an age cut. The vertical distributions and slopes of the three populations in this spatial bin are shown in Fig~\ref{fig:gaia_strlinefit}. There is an offset in the zero point among the different age groups, similar to what is seen in the model. We find a similar explanation as in the model, \ie\ a difference in bending motion amplitudes in different age groups is giving rise to this offset. The presence of a non-zero slope in all three populations indicates that all of them take part in the breathing motion. The values of the best-fit slope vary with stellar age: the (absolute) value of the best-fit slope is largest for the intermediate stars ($2.27 \pm 0.25 \kms$ kpc$^{-1}$) and smallest for the old stars ($1.02 \pm 0.19 \kms$ kpc$^{-1}$) whereas the value of the best-fit slope for the young star falls in between  ($1.61 \pm 0.06 \kms$ kpc$^{-1}$). 
We check whether the stars at the larger distance from the disc mid-plane bias the fits by ignoring the two points at the largest height ($|z| = 1.5 \kpc$). The resulting best-fit slopes ($-1.6 \pm 0.1\kms$ kpc$^{-1}$ for young,  $-2.26 \pm 0.6\kms$ kpc$^{-1}$ for intermediate, and  $-1.5 \pm 0.04\kms$ kpc$^{-1}$ for old stellar populations) match, within the error-bars, with the slopes obtained including the largest heights, indicating that the large $|z|$ points are not biasing the fits.

 \begin{figure*}
       \includegraphics[width=\linewidth]{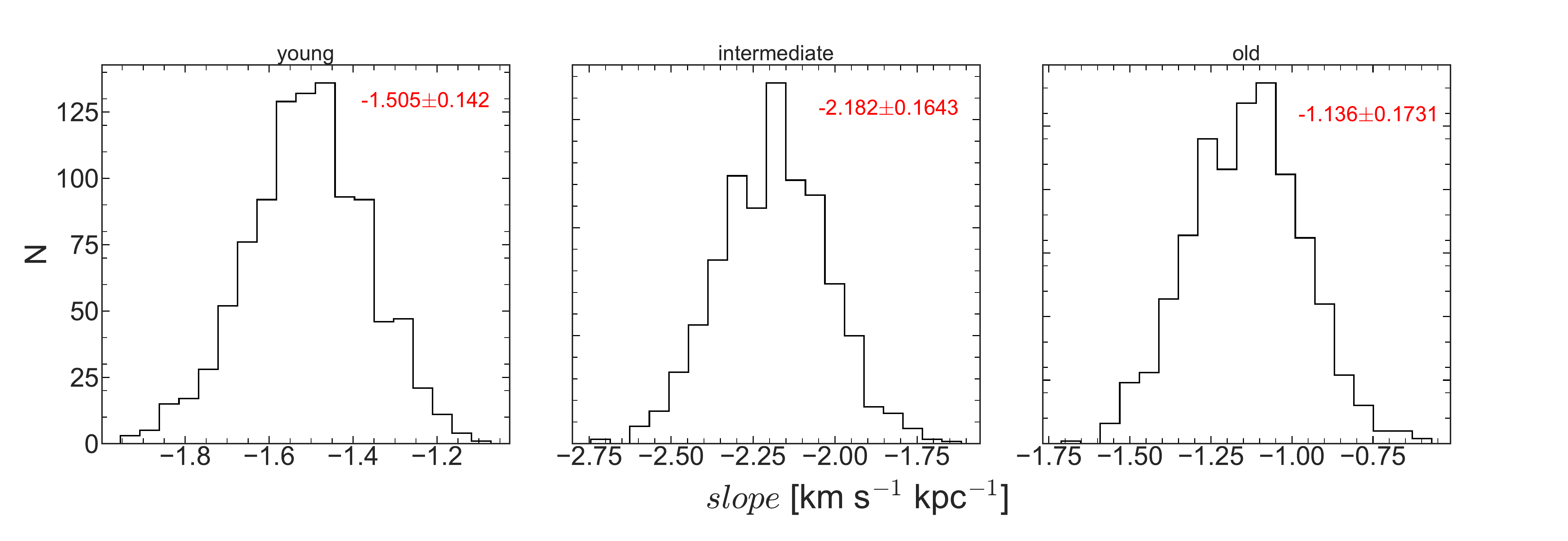}\par
 \caption{SD18 sample: distribution of the best-fit slopes, calculated in the Monte Carlo analysis (for details, see section~\ref{sec:gaiaVertMotion_agecut}) are shown for three stellar populations with different ages, at the spatial location $x \in [7.6, 8.1] \kpc$ and $y \in [0.9, 1.4] \kpc$ . The median values and the associated errors are indicated in each sub-panel.} 
 \label{fig:gaia_strlinefit_errMC}
 \end{figure*}
 
To test the robustness of the best-fit slope to systematic biases in the \gaia\ distances, we lower the distances of all the stars in our selected sub-sample by 10 percent to simulate a parallax bias. After repeating the analysis of fitting a straight line to the variation of $\avg {v_z}$ with height, the resulting change in the best-fit slope is negligible (less than 5 per cent) for all three age populations.  This emphasises that the breathing motions are not sensitive to systematic errors of this type and  accentuating the robustness of the detection of breathing motions in the Solar Neighbourhood.
 
So far, we have estimated the statistical errors on $\avg{v_z}$ and also checked for the systematic errors on the best-fit slope. However, the uncertainties associated with the parallax and the proper motions, the quantities used to derive vertical motion of stars, also result in uncertainties in the calculation of the $\avg{v_z}$ and the corresponding best-fit slope. To estimate this error on the best-fit slope, we perform a Monte Carlo analysis. First, we calculate the slope and the associated error of the best-fit straight line using the 6-D position-velocity of stars in our full selected sub-sample. Since there is a robust signal of breathing motions only in our selected spatial location in the Solar Neighbourhood, we restrict our Monte Carlo analysis for that spatial location only. Then, using the errors in parallax and the proper motion, we add random Gaussian errors in the parallax and the proper motion of each star in our sub-sample, and recalculate the 6-D position-velocity of stars in our selected sub-sample using {\sc Galpy} \citep{Bovy2015}. Once we recalculate the 6-D position-velocity of stars, we repeat the exercise of fitting a straight-line to the variation of $\avg {v_z}$ with height, $z$, for all three stellar populations with different ages. This process is repeated 1000 times. The resulting distributions of best-fit slopes, calculated for that particular spatial bin, for all three stellar populations are shown in Fig.~\ref{fig:gaia_strlinefit_errMC}. The estimated errors in the best-fit slope measurements, obtained from the Monte Carlo analysis are $\sim 14$ per cent for the young population, $\sim 16$ per cent for the intermediate population, and $\sim 17$ per cent for the old population.
 
\section{Signatures of spiral driven breathing motions}
\label{sec:physical_explanation}
%&&&&&&&&&&&&&&&&&&&&&&&&&&&&&&&&&&&&&&&&&&&&&&

Using a star-forming simulation we have shown that a spiral density wave drives vertical breathing motions. The increasing amplitude of the breathing velocity with height above the mid-plane, previously reported by \citet{Debattista2014}, is consistent with the trend seen in the Milky Way \citep{GaiaCollaboration2018}. This behaviour is an important clue to the source of breathing motions: it must be something local to affect stars near the mid-plane differently from those further up.  The reason why spirals have this effect is easy to understand: on vertical scales small compared to the disc scale-length, the vertical gravitational field can be approximated, using Gauss's law, as a uniform field from an infinite sheet of mass corresponding to the mass enclosed within a cylinder of the same height and small radius.  Both the simulation of \citet{Debattista2014} and the one considered here show that the spiral density wave extends to large heights (see Fig.~\ref{fig:residualmap_phiz}).  If the density of the disc is decomposed into an axisymmetric part and a spiral perturbation part:
\begin{equation}
\rho(R,\phi,z) = \rho_0(R,z) + \rho_S(R,\phi,z),
\end{equation}
then the vertical velocities are in equilibrium with the axisymmetric part of the potential and $\rho_0(R,z)$ gives rise to $\vb = 0$. Since the azimuthally varying part, $\rho_S$, is due to spirals, it extends to large height. The enclosed mass therefore increases with height as the vertical integral of $\rho_S(R,\phi,z)$.  Therefore the gravitational field increases with distance from the mid-plane, and \vb\ increases with height, as we see in the simulations, and in the Milky Way (compare Fig.~\ref{fig:breath_xymap_nbodysph} here and Fig. C.6 in \citet{Katzetal2018}).  It is noteworthy that this vertical increase in \vb\ is absent in the model of \citet{Faureetal2014}, who used an analytic spiral potential with a very short scale height of $100\pc$, much shorter than that of the unperturbed density distribution.  As a result, the gravitational field reaches nearly constant value quite close to the mid-plane in their model. Consequently, \vb\ shows a near-linear increase from the disc mid-plane, and after $\sim 0.7 z_0$, it starts to decline where $z_0$ is the scale-height of the spiral perturbation. The vertical variation of \vb\ in the \gaia\ data therefore indicates that the spiral structure in the Milky Way is vertically extended.
% %
The vertical increase of \vb\ is different from the nearly constant \vbend\ as a function of height (compare both panels of Fig. C.6 in \citet{Katzetal2018}). This indicates that the Milky Way's bending motions are probably not excited by spirals. 

Moreover we examined the dependence of the breathing velocity on stellar age. The star-forming model exhibits a decreasing \vb\ with increasing stellar age (see Fig.~\ref{fig:compare_slopevsazimuth_sim}) indicating that increasing random motions decrease the vertical response of the stars. Fig.~\ref{fig:gaia_strlinefit} shows that, using the SD18 dataset, the intermediate stellar population indeed yields a larger (absolute) value of the best-fit slope in the Solar Neighbourhood compared with the old stellar population, while the young stellar population has a slope that falls in between;  however the difference between the young and intermediate age populations ($1.8 \sigma$) is not statistically significant in this dataset. In other words, the breathing motion is stronger for the younger stellar populations (i.e., young and intermediate populations considered together) than that for the old stellar population in the Solar Neighbourhood.

For these reasons we therefore argue that the breathing motions seen in the Solar Neighbourhood are very likely to be driven by the spiral density wave structure. \citet{Widrowetal2014} proposed a scenario where a satellite galaxy, while plunging into the disc, can generate both bending and breathing motions, depending on the value of the satellite's velocity in the direction perpendicular to the disc. Such interactions also excite a strong spiral response within the disc which are probably partly responsible for the breathing motions seen in the simulations of \citet{Widrowetal2014}.  The vertical variation of the amplitude of the breathing modes excited by satellites is also unclear.

\section{Future prospects and summary}
\label{sec:discussion}
%&&&&&&&&&&&&&&&&&&&&&&&&&&&&&&&&&&&&&&&&&&&&&&&&&&&&&&&&&&&&&&&&&&&&&&&&&&&

Our requirement of stellar ages limits the number of stars available to us by almost a factor of two when compared with the \gaia\ {\sc main sample}  \citep[which contains 6,376,803 stars; ][]{Katzetal2018}. In addition, the azimuthal coverage in \gaia\ DR2 is not large ($\Delta \phi \sim 30\degrees$). This, in turn, prevents us from exploring the change from expanding to compressing breathing motions as a function of azimuthal angle, which would help constrain the extent of the spiral and its corotation radius \citep{Faureetal2014, Debattista2014}. \gaia\ DR3 holds the potential to permit us to investigate the variation of breathing motions as a function of azimuth, in order to constrain the Milky Way's spiral structure better.

Our main findings are:\\

\begin{itemize}
\item{The spirals in the self-consistent star-forming model drive coherent vertical breathing motions, with amplitude which increases with the height from the disc mid-plane. We argue that this vertical variation is a consequence of the vertically extended spiral structure.  Since \gaia\ DR2 also exhibits this increasing breathing amplitude with height, it implies that the Milky Way's spirals are also vertically extended.}
\item{Prominent breathing motions are present in stellar populations of all ages in the model, with the strongest motions in the young stellar population, and weakest in the old one.}
\item{Using \gaia\ DR2 with complementary age data from \citet{SandersandDas2018}, we showed that the location near the Solar Neighbourhood with the smallest biases shows significant, strong breathing motions for all ages. In addition, the amplitude of the breathing motions varies with stellar age, with the oldest population displaying the weakest breathing amplitude (when compared with the intermediate and the young population, at $2.8-3.1\sigma$). The difference between the young and intermediate age populations is smaller but less statistically significant ($1.8\sigma$). This behaviour is similar to the trend shown in the simulation model and is consistent with spirals being the driving mechanism of breathing modes.}
\end{itemize}

The resemblance in the height and age variation of the amplitude of breathing velocity, in the star-forming model and in the location in the Milky Way showing prominent breathing motions from the \gaia\ DR2 dataset indicates that the observed breathing motions in the Milky Way are likely to be excited by spiral density waves. Due to the limited coverage and uncertainties in the \gaia\ parallax, we have restricted to the Solar Neighbourhood our search for breathing motions, but with the upcoming \gaia\ DR3, there will be a larger number of stars allowing us to improve the statistics in the Solar Neighbourhood.

\section*{Acknowledgement}

 We thank the anonymous referee for useful comments which helped to improve this paper.
S.G. acknowledges support from an Indo-French CEFIPRA project (Project No.: 5804-1). 
V.P.D. is supported by STFC Consolidated grant ST/R000786/1. The simulation in this paper was run at the DiRAC Shared Memory Processing system at the University of Cambridge, operated by the COSMOS Project at the Department of Applied Mathematics and Theoretical Physics on behalf of the STFC DiRAC HPC Facility (\href {www.dirac.ac.uk}{www.dirac.ac.uk}). This equipment was funded by BIS National E-infrastructure capital grant ST/J005673/1, STFC capital grant ST/H008586/1, and STFC DiRAC Operations grant ST/K00333X/1. DiRAC is part of the National E-Infrastructure.  

\section*{Data availability}
The simulation data underlying this article will be shared on reasonable request to V.P.D (vpdebattista@gmail.com). The Sanders \& Das dataset is publicly available from \href {https://www.ast.cam.ac.uk/~jls/data/gaia_spectro.hdf5}{this URL}.

\bibliography{my_ref}{}

\begin{thebibliography}{}
\makeatletter
\relax
\def\mn@urlcharsother{\let\do\@makeother \do\$\do\&\do\#\do\^\do\_\do\%\do\~}
\def\mn@doi{\begingroup\mn@urlcharsother \@ifnextchar [ {\mn@doi@}
  {\mn@doi@[]}}
\def\mn@doi@[#1]#2{\def\@tempa{#1}\ifx\@tempa\@empty \href
  {http://dx.doi.org/#2} {doi:#2}\else \href {http://dx.doi.org/#2} {#1}\fi
  \endgroup}
\def\mn@eprint#1#2{\mn@eprint@#1:#2::\@nil}
\def\mn@eprint@arXiv#1{\href {http://arxiv.org/abs/#1} {{\tt arXiv:#1}}}
\def\mn@eprint@dblp#1{\href {http://dblp.uni-trier.de/rec/bibtex/#1.xml}
  {dblp:#1}}
\def\mn@eprint@#1:#2:#3:#4\@nil{\def\@tempa {#1}\def\@tempb {#2}\def\@tempc
  {#3}\ifx \@tempc \@empty \let \@tempc \@tempb \let \@tempb \@tempa \fi \ifx
  \@tempb \@empty \def\@tempb {arXiv}\fi \@ifundefined
  {mn@eprint@\@tempb}{\@tempb:\@tempc}{\expandafter \expandafter \csname
  mn@eprint@\@tempb\endcsname \expandafter{\@tempc}}}

\bibitem[\protect\citeauthoryear{{Antoja} et~al.,}{{Antoja}
  et~al.}{2018}]{Antojaetal2018}
{Antoja} T.,  et~al., 2018, \mn@doi [\nat] {10.1038/s41586-018-0510-7}, \href
  {http://adsabs.harvard.edu/abs/2018Natur.561..360A} {561, 360}

\bibitem[\protect\citeauthoryear{{Araki}}{{Araki}}{1985}]{Araki1985}
{Araki} S.,  1985, PhD thesis, MASSACHUSETTS INSTITUTE OF TECHNOLOGY.

\bibitem[\protect\citeauthoryear{{Athanassoula}}{{Athanassoula}}{2012}]{Athanassoula2012}
{Athanassoula} E.,  2012, \mn@doi [\mnras] {10.1111/j.1745-3933.2012.01320.x},
  \href {https://ui.adsabs.harvard.edu/abs/2012MNRAS.426L..46A} {426, L46}

\bibitem[\protect\citeauthoryear{{Athanassoula}, {Romero-G{\'o}mez}  \&
  {Masdemont}}{{Athanassoula} et~al.}{2009}]{Athanassoulaetal2009}
{Athanassoula} E.,  {Romero-G{\'o}mez} M.,   {Masdemont} J.~J.,  2009, \mn@doi
  [\mnras] {10.1111/j.1365-2966.2008.14273.x}, \href
  {https://ui.adsabs.harvard.edu/abs/2009MNRAS.394...67A} {394, 67}

\bibitem[\protect\citeauthoryear{{Athanassoula}, {Romero-G{\'o}mez}, {Bosma}
  \& {Masdemont}}{{Athanassoula} et~al.}{2010}]{Athanassoulaetal2010}
{Athanassoula} E.,  {Romero-G{\'o}mez} M.,  {Bosma} A.,   {Masdemont} J.~J.,
  2010, \mn@doi [\mnras] {10.1111/j.1365-2966.2010.17010.x}, \href
  {https://ui.adsabs.harvard.edu/abs/2010MNRAS.407.1433A} {407, 1433}

\bibitem[\protect\citeauthoryear{Bailer-Jones, Rybizki, Fouesneau, Mantelet  \&
  Andrae}{Bailer-Jones et~al.}{2018}]{Bailer_Jones_2018}
Bailer-Jones C. A.~L.,  Rybizki J.,  Fouesneau M.,  Mantelet G.,   Andrae R.,
  2018, \mn@doi [The Astronomical Journal] {10.3847/1538-3881/aacb21}, 156, 58

\bibitem[\protect\citeauthoryear{{Bennett} \& {Bovy}}{{Bennett} \&
  {Bovy}}{2019}]{BennettandBovy2019}
{Bennett} M.,  {Bovy} J.,  2019, \mn@doi [\mnras] {10.1093/mnras/sty2813},
  \href {https://ui.adsabs.harvard.edu/abs/2019MNRAS.482.1417B} {482, 1417}

\bibitem[\protect\citeauthoryear{{Binney}}{{Binney}}{2012}]{Binney12}
{Binney} J.,  2012, \mn@doi [\mnras] {10.1111/j.1365-2966.2012.21757.x}, \href
  {https://ui.adsabs.harvard.edu/abs/2012MNRAS.426.1324B} {426, 1324}

\bibitem[\protect\citeauthoryear{{Binney} \& {Tremaine}}{{Binney} \&
  {Tremaine}}{2008}]{BT08}
{Binney} J.,  {Tremaine} S.,  2008, {Galactic Dynamics: Second Edition}.
Princeton University Press

\bibitem[\protect\citeauthoryear{{Bland-Hawthorn} \&
  {Gerhard}}{{Bland-Hawthorn} \& {Gerhard}}{2016}]{Bland-Hawthorn2016}
{Bland-Hawthorn} J.,  {Gerhard} O.,  2016, \mn@doi [\araa]
  {10.1146/annurev-astro-081915-023441}, \href
  {https://ui.adsabs.harvard.edu/abs/2016ARA&A..54..529B} {54, 529}

\bibitem[\protect\citeauthoryear{{Bovy}}{{Bovy}}{2015}]{Bovy2015}
{Bovy} J.,  2015, \mn@doi [\apjs] {10.1088/0067-0049/216/2/29}, \href
  {https://ui.adsabs.harvard.edu/abs/2015ApJS..216...29B} {216, 29}

\bibitem[\protect\citeauthoryear{{Bullock}, {Dekel}, {Kolatt}, {Kravtsov},
  {Klypin}, {Porciani}  \& {Primack}}{{Bullock} et~al.}{2001}]{Bullocketal2001}
{Bullock} J.~S.,  {Dekel} A.,  {Kolatt} T.~S.,  {Kravtsov} A.~V.,  {Klypin}
  A.~A.,  {Porciani} C.,   {Primack} J.~R.,  2001, \mn@doi [\apj]
  {10.1086/321477}, \href
  {https://ui.adsabs.harvard.edu/abs/2001ApJ...555..240B} {555, 240}

\bibitem[\protect\citeauthoryear{{Carlin} et~al.,}{{Carlin}
  et~al.}{2013}]{Carlinetal2013}
{Carlin} J.~L.,  et~al., 2013, \mn@doi [\apjl] {10.1088/2041-8205/777/1/L5},
  \href {https://ui.adsabs.harvard.edu/abs/2013ApJ...777L...5C} {777, L5}

\bibitem[\protect\citeauthoryear{{Carrillo} et~al.,}{{Carrillo}
  et~al.}{2018}]{Carrilloetal2018}
{Carrillo} I.,  et~al., 2018, \mn@doi [\mnras] {10.1093/mnras/stx3342}, \href
  {https://ui.adsabs.harvard.edu/abs/2018MNRAS.475.2679C} {475, 2679}

\bibitem[\protect\citeauthoryear{{Chequers}, {Widrow}  \& {Darling}}{{Chequers}
  et~al.}{2018}]{Chequersetal2018}
{Chequers} M.~H.,  {Widrow} L.~M.,   {Darling} K.,  2018, \mn@doi [\mnras]
  {10.1093/mnras/sty2114}, \href
  {https://ui.adsabs.harvard.edu/abs/2018MNRAS.480.4244C} {480, 4244}

\bibitem[\protect\citeauthoryear{{D'Onghia}, {Vogelsberger}  \&
  {Hernquist}}{{D'Onghia} et~al.}{2013}]{Donghia2013}
{D'Onghia} E.,  {Vogelsberger} M.,   {Hernquist} L.,  2013, \mn@doi [\apj]
  {10.1088/0004-637X/766/1/34}, \href
  {http://adsabs.harvard.edu/abs/2013ApJ...766...34D} {766, 34}

\bibitem[\protect\citeauthoryear{{D'Onghia}, {Madau}, {Vera-Ciro}, {Quillen}
  \& {Hernquist}}{{D'Onghia} et~al.}{2016}]{Donghiaetal2016}
{D'Onghia} E.,  {Madau} P.,  {Vera-Ciro} C.,  {Quillen} A.,   {Hernquist} L.,
  2016, \mn@doi [\apj] {10.3847/0004-637X/823/1/4}, \href
  {https://ui.adsabs.harvard.edu/abs/2016ApJ...823....4D} {823, 4}

\bibitem[\protect\citeauthoryear{{Debattista}}{{Debattista}}{2014}]{Debattista2014}
{Debattista} V.~P.,  2014, \mn@doi [\mnras] {10.1093/mnrasl/slu069}, \href
  {https://ui.adsabs.harvard.edu/abs/2014MNRAS.443L...1D} {443, L1}

\bibitem[\protect\citeauthoryear{{Dobbs} \& {Baba}}{{Dobbs} \&
  {Baba}}{2014}]{DobbsandBaba2014}
{Dobbs} C.,  {Baba} J.,  2014, \mn@doi [\pasa] {10.1017/pasa.2014.31}, \href
  {http://adsabs.harvard.edu/abs/2014PASA...31...35D} {31, e035}

\bibitem[\protect\citeauthoryear{{Dobbs}, {Theis}, {Pringle}  \&
  {Bate}}{{Dobbs} et~al.}{2010}]{Dobbsetal2010}
{Dobbs} C.~L.,  {Theis} C.,  {Pringle} J.~E.,   {Bate} M.~R.,  2010, \mn@doi
  [\mnras] {10.1111/j.1365-2966.2009.16161.x}, \href
  {https://ui.adsabs.harvard.edu/abs/2010MNRAS.403..625D} {403, 625}

\bibitem[\protect\citeauthoryear{{Efthymiopoulos}, {Harsoula}  \&
  {Contopoulos}}{{Efthymiopoulos} et~al.}{2020}]{Efthymiopoulosetal2020}
{Efthymiopoulos} C.,  {Harsoula} M.,   {Contopoulos} G.,  2020, \mn@doi [\aap]
  {10.1051/0004-6361/201936871}, \href
  {https://ui.adsabs.harvard.edu/abs/2020A&A...636A..44E} {636, A44}

\bibitem[\protect\citeauthoryear{{Elmegreen} \& {Elmegreen}}{{Elmegreen} \&
  {Elmegreen}}{2014}]{Elmegreenetal2014}
{Elmegreen} D.~M.,  {Elmegreen} B.~G.,  2014, \mn@doi [\apj]
  {10.1088/0004-637X/781/1/11}, \href
  {http://adsabs.harvard.edu/abs/2014ApJ...781...11E} {781, 11}

\bibitem[\protect\citeauthoryear{{Elmegreen} et~al.,}{{Elmegreen}
  et~al.}{2011}]{Elmegreenetal2011}
{Elmegreen} D.~M.,  et~al., 2011, \mn@doi [\apj] {10.1088/0004-637X/737/1/32},
  \href {http://adsabs.harvard.edu/abs/2011ApJ...737...32E} {737, 32}

\bibitem[\protect\citeauthoryear{{Faure}, {Siebert}  \& {Famaey}}{{Faure}
  et~al.}{2014}]{Faureetal2014}
{Faure} C.,  {Siebert} A.,   {Famaey} B.,  2014, \mn@doi [\mnras]
  {10.1093/mnras/stu428}, \href
  {https://ui.adsabs.harvard.edu/abs/2014MNRAS.440.2564F} {440, 2564}

\bibitem[\protect\citeauthoryear{{Gaia Collaboration} et~al.,}{{Gaia
  Collaboration} et~al.}{2016}]{GaiaCollaboration2016}
{Gaia Collaboration} et~al., 2016, \mn@doi [\aap]
  {10.1051/0004-6361/201629272}, \href
  {https://ui.adsabs.harvard.edu/abs/2016A%26A...595A...1G} {595, A1}

\bibitem[\protect\citeauthoryear{{Gaia Collaboration} et~al.,}{{Gaia
  Collaboration} et~al.}{2018a}]{GaiaCollaboration2018}
{Gaia Collaboration} et~al., 2018a, \mn@doi [\aap]
  {10.1051/0004-6361/201833051}, \href
  {https://ui.adsabs.harvard.edu/abs/2018A%26A...616A...1G} {616, A1}

\bibitem[\protect\citeauthoryear{{Gaia Collaboration} et~al.,}{{Gaia
  Collaboration} et~al.}{2018b}]{Katzetal2018}
{Gaia Collaboration} et~al., 2018b, \mn@doi [\aap]
  {10.1051/0004-6361/201832865}, \href
  {http://adsabs.harvard.edu/abs/2018A%26A...616A..11G} {616, A11}

\bibitem[\protect\citeauthoryear{{Gerhard}}{{Gerhard}}{2002}]{Gerhard2002}
{Gerhard} O.,  2002, in {Da Costa} G.~S.,  {Sadler} E.~M.,   {Jerjen} H.,  eds,
   Astronomical Society of the Pacific Conference Series Vol. 273, The
  Dynamics, Structure \& History of Galaxies: A Workshop in Honour of Professor
  Ken Freeman. p.~73 (\mn@eprint {arXiv} {astro-ph/0203109})

\bibitem[\protect\citeauthoryear{{Ghosh} \& {Jog}}{{Ghosh} \&
  {Jog}}{2015}]{GhoshJog2015}
{Ghosh} S.,  {Jog} C.~J.,  2015, \mn@doi [\mnras] {10.1093/mnras/stv1040},
  \href {http://adsabs.harvard.edu/abs/2015MNRAS.451.1350G} {451, 1350}

\bibitem[\protect\citeauthoryear{{Ghosh} \& {Jog}}{{Ghosh} \&
  {Jog}}{2016}]{GhoshJog2016}
{Ghosh} S.,  {Jog} C.~J.,  2016, \mn@doi [\mnras] {10.1093/mnras/stw914}, \href
  {http://adsabs.harvard.edu/abs/2016MNRAS.459.4057G} {459, 4057}

\bibitem[\protect\citeauthoryear{{Goldreich} \& {Lynden-Bell}}{{Goldreich} \&
  {Lynden-Bell}}{1965}]{GoldreichLyden65}
{Goldreich} P.,  {Lynden-Bell} D.,  1965, \mn@doi [\mnras]
  {10.1093/mnras/130.2.125}, \href
  {http://adsabs.harvard.edu/abs/1965MNRAS.130..125G} {130, 125}

\bibitem[\protect\citeauthoryear{{G{\'o}mez}, {Minchev}, {O'Shea}, {Beers},
  {Bullock}  \& {Purcell}}{{G{\'o}mez} et~al.}{2013}]{Gomezetal2013}
{G{\'o}mez} F.~A.,  {Minchev} I.,  {O'Shea} B.~W.,  {Beers} T.~C.,  {Bullock}
  J.~S.,   {Purcell} C.~W.,  2013, \mn@doi [\mnras] {10.1093/mnras/sts327},
  \href {https://ui.adsabs.harvard.edu/abs/2013MNRAS.429..159G} {429, 159}

\bibitem[\protect\citeauthoryear{{Hodge} et~al.,}{{Hodge}
  et~al.}{2019}]{Hodgeetal2019}
{Hodge} J.~A.,  et~al., 2019, \mn@doi [\apj] {10.3847/1538-4357/ab1846}, \href
  {https://ui.adsabs.harvard.edu/abs/2019ApJ...876..130H} {876, 130}

\bibitem[\protect\citeauthoryear{{Hunter} \& {Toomre}}{{Hunter} \&
  {Toomre}}{1969}]{HunterandToomre1969}
{Hunter} C.,  {Toomre} A.,  1969, \mn@doi [\apj] {10.1086/149908}, \href
  {https://ui.adsabs.harvard.edu/abs/1969ApJ...155..747H} {155, 747}

\bibitem[\protect\citeauthoryear{{Julian} \& {Toomre}}{{Julian} \&
  {Toomre}}{1966}]{JulainToomre66}
{Julian} W.~H.,  {Toomre} A.,  1966, \mn@doi [\apj] {10.1086/148957}, \href
  {http://adsabs.harvard.edu/abs/1966ApJ...146..810J} {146, 810}

\bibitem[\protect\citeauthoryear{{Khoperskov}, {Di Matteo}, {Gerhard}, {Katz},
  {Haywood}, {Combes}, {Berczik}  \& {Gomez}}{{Khoperskov}
  et~al.}{2019}]{khoperskov2019}
{Khoperskov} S.,  {Di Matteo} P.,  {Gerhard} O.,  {Katz} D.,  {Haywood} M.,
  {Combes} F.,  {Berczik} P.,   {Gomez} A.,  2019, \mn@doi [\aap]
  {10.1051/0004-6361/201834707}, \href
  {https://ui.adsabs.harvard.edu/abs/2019A&A...622L...6K} {622, L6}

\bibitem[\protect\citeauthoryear{{Lindegren} et~al.,}{{Lindegren}
  et~al.}{2018}]{Lindegrenetal2018}
{Lindegren} L.,  et~al., 2018, \mn@doi [\aap] {10.1051/0004-6361/201832727},
  \href {https://ui.adsabs.harvard.edu/abs/2018A&A...616A...2L} {616, A2}

\bibitem[\protect\citeauthoryear{{Loebman}, {Debattista}, {Nidever}, {Hayden},
  {Holtzman}, {Clarke}, {Ro{\v{s}}kar}  \& {Valluri}}{{Loebman}
  et~al.}{2016}]{Loebmanetal2016}
{Loebman} S.~R.,  {Debattista} V.~P.,  {Nidever} D.~L.,  {Hayden} M.~R.,
  {Holtzman} J.~A.,  {Clarke} A.~J.,  {Ro{\v{s}}kar} R.,   {Valluri} M.,  2016,
  \mn@doi [\apjl] {10.3847/2041-8205/818/1/L6}, \href
  {https://ui.adsabs.harvard.edu/abs/2016ApJ...818L...6L} {818, L6}

\bibitem[\protect\citeauthoryear{{Luri} et~al.,}{{Luri}
  et~al.}{2018}]{Lurietal2018}
{Luri} X.,  et~al., 2018, \mn@doi [\aap] {10.1051/0004-6361/201832964}, \href
  {https://ui.adsabs.harvard.edu/abs/2018A&A...616A...9L} {616, A9}

\bibitem[\protect\citeauthoryear{{Lynden-Bell} \& {Kalnajs}}{{Lynden-Bell} \&
  {Kalnajs}}{1972}]{LyndenBellKalnajs1972}
{Lynden-Bell} D.,  {Kalnajs} A.~J.,  1972, \mn@doi [\mnras]
  {10.1093/mnras/157.1.1}, \href
  {https://ui.adsabs.harvard.edu/abs/1972MNRAS.157....1L} {157, 1}

\bibitem[\protect\citeauthoryear{{Masset} \& {Tagger}}{{Masset} \&
  {Tagger}}{1997}]{Masset1997}
{Masset} F.,  {Tagger} M.,  1997, \aap, \href
  {http://adsabs.harvard.edu/abs/1997A%26A...322..442M} {322, 442}

\bibitem[\protect\citeauthoryear{{Mathur}}{{Mathur}}{1990}]{Mathur1990}
{Mathur} S.~D.,  1990, \mnras, \href
  {https://ui.adsabs.harvard.edu/abs/1990MNRAS.243..529M} {243, 529}

\bibitem[\protect\citeauthoryear{{McMillan}}{{McMillan}}{2017}]{McMillan}
{McMillan} P.~J.,  2017, \mn@doi [\mnras] {10.1093/mnras/stw2759}, \href
  {https://ui.adsabs.harvard.edu/abs/2017MNRAS.465...76M} {465, 76}

\bibitem[\protect\citeauthoryear{{Monari}, {Famaey}  \& {Siebert}}{{Monari}
  et~al.}{2015}]{Monarietal2015}
{Monari} G.,  {Famaey} B.,   {Siebert} A.,  2015, \mn@doi [\mnras]
  {10.1093/mnras/stv1206}, \href
  {https://ui.adsabs.harvard.edu/abs/2015MNRAS.452..747M} {452, 747}

\bibitem[\protect\citeauthoryear{{Monari}, {Famaey}  \& {Siebert}}{{Monari}
  et~al.}{2016}]{Monarietal2016}
{Monari} G.,  {Famaey} B.,   {Siebert} A.,  2016, \mn@doi [\mnras]
  {10.1093/mnras/stw171}, \href
  {https://ui.adsabs.harvard.edu/abs/2016MNRAS.457.2569M} {457, 2569}

\bibitem[\protect\citeauthoryear{{Navarro}, {Frenk}  \& {White}}{{Navarro}
  et~al.}{1996}]{nfw}
{Navarro} J.~F.,  {Frenk} C.~S.,   {White} S. D.~M.,  1996, \mn@doi [\apj]
  {10.1086/177173}, \href
  {https://ui.adsabs.harvard.edu/abs/1996ApJ...462..563N} {462, 563}

\bibitem[\protect\citeauthoryear{{Rix} \& {Zaritsky}}{{Rix} \&
  {Zaritsky}}{1995}]{Rix1995}
{Rix} H.-W.,  {Zaritsky} D.,  1995, \mn@doi [\apj] {10.1086/175858}, \href
  {https://ui.adsabs.harvard.edu/abs/1995ApJ...447...82R} {447, 82}

\bibitem[\protect\citeauthoryear{{Ro{\v{s}}kar}, {Debattista}, {Quinn},
  {Stinson}  \& {Wadsley}}{{Ro{\v{s}}kar} et~al.}{2008}]{Roskaretal2008}
{Ro{\v{s}}kar} R.,  {Debattista} V.~P.,  {Quinn} T.~R.,  {Stinson} G.~S.,
  {Wadsley} J.,  2008, \mn@doi [\apjl] {10.1086/592231}, \href
  {https://ui.adsabs.harvard.edu/abs/2008ApJ...684L..79R} {684, L79}

\bibitem[\protect\citeauthoryear{{Ro{\v{s}}kar}, {Debattista}, {Quinn}  \&
  {Wadsley}}{{Ro{\v{s}}kar} et~al.}{2012}]{Roskaretal2012}
{Ro{\v{s}}kar} R.,  {Debattista} V.~P.,  {Quinn} T.~R.,   {Wadsley} J.,  2012,
  \mn@doi [\mnras] {10.1111/j.1365-2966.2012.21860.x}, \href
  {https://ui.adsabs.harvard.edu/abs/2012MNRAS.426.2089R} {426, 2089}

\bibitem[\protect\citeauthoryear{{Salo}, {Laurikainen}, {Buta}  \&
  {Knapen}}{{Salo} et~al.}{2010}]{Salo2010}
{Salo} H.,  {Laurikainen} E.,  {Buta} R.,   {Knapen} J.~H.,  2010, \mn@doi
  [\apjl] {10.1088/2041-8205/715/1/L56}, \href
  {http://adsabs.harvard.edu/abs/2010ApJ...715L..56S} {715, L56}

\bibitem[\protect\citeauthoryear{{Sanders} \& {Binney}}{{Sanders} \&
  {Binney}}{2016}]{Sanders+16}
{Sanders} J.~L.,  {Binney} J.,  2016, \mn@doi [\mnras] {10.1093/mnras/stw106},
  \href {https://ui.adsabs.harvard.edu/abs/2016MNRAS.457.2107S} {457, 2107}

\bibitem[\protect\citeauthoryear{{Sanders} \& {Das}}{{Sanders} \&
  {Das}}{2018}]{SandersandDas2018}
{Sanders} J.~L.,  {Das} P.,  2018, \mn@doi [\mnras] {10.1093/mnras/sty2490},
  \href {https://ui.adsabs.harvard.edu/abs/2018MNRAS.481.4093S} {481, 4093}

\bibitem[\protect\citeauthoryear{{Savchenko}, {Marchuk}, {Mosenkov}  \&
  {Grishunin}}{{Savchenko} et~al.}{2020}]{Savchenkoetal2020}
{Savchenko} S.,  {Marchuk} A.,  {Mosenkov} A.,   {Grishunin} K.,  2020, \mn@doi
  [\mnras] {10.1093/mnras/staa258}, \href
  {https://ui.adsabs.harvard.edu/abs/2020MNRAS.493..390S} {493, 390}

\bibitem[\protect\citeauthoryear{{Sch{\"o}nrich} \& {Binney}}{{Sch{\"o}nrich}
  \& {Binney}}{2009}]{SchonrichBinney2009}
{Sch{\"o}nrich} R.,  {Binney} J.,  2009, \mn@doi [\mnras]
  {10.1111/j.1365-2966.2009.14750.x}, \href
  {https://ui.adsabs.harvard.edu/abs/2009MNRAS.396..203S} {396, 203}

\bibitem[\protect\citeauthoryear{{Sch{\"o}nrich}, {Binney}  \&
  {Dehnen}}{{Sch{\"o}nrich} et~al.}{2010}]{schronrichetal2010}
{Sch{\"o}nrich} R.,  {Binney} J.,   {Dehnen} W.,  2010, \mn@doi [\mnras]
  {10.1111/j.1365-2966.2010.16253.x}, \href
  {https://ui.adsabs.harvard.edu/abs/2010MNRAS.403.1829S} {403, 1829}

\bibitem[\protect\citeauthoryear{{Sch{\"o}nrich}, {McMillan}  \&
  {Eyer}}{{Sch{\"o}nrich} et~al.}{2019}]{Schoenrichetal2019}
{Sch{\"o}nrich} R.,  {McMillan} P.,   {Eyer} L.,  2019, \mn@doi [\mnras]
  {10.1093/mnras/stz1451}, \href
  {https://ui.adsabs.harvard.edu/abs/2019MNRAS.487.3568S} {487, 3568}

\bibitem[\protect\citeauthoryear{{Sellwood}}{{Sellwood}}{2012}]{Sellwood2012}
{Sellwood} J.~A.,  2012, \mn@doi [\apj] {10.1088/0004-637X/751/1/44}, \href
  {https://ui.adsabs.harvard.edu/abs/2012ApJ...751...44S} {751, 44}

\bibitem[\protect\citeauthoryear{{Sellwood} \& {Binney}}{{Sellwood} \&
  {Binney}}{2002}]{SellwoodBinney2002}
{Sellwood} J.~A.,  {Binney} J.~J.,  2002, \mn@doi [\mnras]
  {10.1046/j.1365-8711.2002.05806.x}, \href
  {https://ui.adsabs.harvard.edu/abs/2002MNRAS.336..785S} {336, 785}

\bibitem[\protect\citeauthoryear{{Sellwood} \& {Carlberg}}{{Sellwood} \&
  {Carlberg}}{1984}]{SellwoodCarlberg1984}
{Sellwood} J.~A.,  {Carlberg} R.~G.,  1984, \mn@doi [\apj] {10.1086/162176},
  \href {https://ui.adsabs.harvard.edu/abs/1984ApJ...282...61S} {282, 61}

\bibitem[\protect\citeauthoryear{{Sellwood} \& {Carlberg}}{{Sellwood} \&
  {Carlberg}}{2019}]{SellwoodandCarlberg2019}
{Sellwood} J.~A.,  {Carlberg} R.~G.,  2019, \mn@doi [\mnras]
  {10.1093/mnras/stz2132}, \href
  {https://ui.adsabs.harvard.edu/abs/2019MNRAS.489..116S} {489, 116}

\bibitem[\protect\citeauthoryear{{Sellwood} \& {Kahn}}{{Sellwood} \&
  {Kahn}}{1991}]{SellwoodKahn1991}
{Sellwood} J.~A.,  {Kahn} F.~D.,  1991, \mn@doi [\mnras]
  {10.1093/mnras/250.2.278}, \href
  {https://ui.adsabs.harvard.edu/abs/1991MNRAS.250..278S} {250, 278}

\bibitem[\protect\citeauthoryear{{Sellwood} \& {Lin}}{{Sellwood} \&
  {Lin}}{1989}]{SellwoodLin1989}
{Sellwood} J.~A.,  {Lin} D.~N.~C.,  1989, \mn@doi [\mnras]
  {10.1093/mnras/240.4.991}, \href
  {https://ui.adsabs.harvard.edu/abs/1989MNRAS.240..991S} {240, 991}

\bibitem[\protect\citeauthoryear{{Shen}, {Rich}, {Kormendy}, {Howard}, {De
  Propris}  \& {Kunder}}{{Shen} et~al.}{2010}]{Shenetal2010}
{Shen} J.,  {Rich} R.~M.,  {Kormendy} J.,  {Howard} C.~D.,  {De Propris} R.,
  {Kunder} A.,  2010, \mn@doi [\apjl] {10.1088/2041-8205/720/1/L72}, \href
  {https://ui.adsabs.harvard.edu/abs/2010ApJ...720L..72S} {720, L72}

\bibitem[\protect\citeauthoryear{{Siebert} et~al.,}{{Siebert}
  et~al.}{2011}]{Siebertetal2011}
{Siebert} A.,  et~al., 2011, \mn@doi [\mnras]
  {10.1111/j.1365-2966.2010.18037.x}, \href
  {https://ui.adsabs.harvard.edu/abs/2011MNRAS.412.2026S} {412, 2026}

\bibitem[\protect\citeauthoryear{{Siebert} et~al.,}{{Siebert}
  et~al.}{2012}]{Siebertetal2012}
{Siebert} A.,  et~al., 2012, \mn@doi [\mnras]
  {10.1111/j.1365-2966.2012.21638.x}, \href
  {https://ui.adsabs.harvard.edu/abs/2012MNRAS.425.2335S} {425, 2335}

\bibitem[\protect\citeauthoryear{{Stinson}, {Seth}, {Katz}, {Wadsley},
  {Governato}  \& {Quinn}}{{Stinson} et~al.}{2006}]{Stinsonetal2006}
{Stinson} G.,  {Seth} A.,  {Katz} N.,  {Wadsley} J.,  {Governato} F.,   {Quinn}
  T.,  2006, \mn@doi [\mnras] {10.1111/j.1365-2966.2006.11097.x}, \href
  {https://ui.adsabs.harvard.edu/abs/2006MNRAS.373.1074S} {373, 1074}

\bibitem[\protect\citeauthoryear{{Toomre}}{{Toomre}}{1981}]{Toomre81}
{Toomre} A.,  1981, in {Fall} S.~M.,  {Lynden-Bell} D.,  eds, Structure and
  Evolution of Normal Galaxies. pp 111--136

\bibitem[\protect\citeauthoryear{{Toomre} \& {Toomre}}{{Toomre} \&
  {Toomre}}{1972}]{Toomre1972}
{Toomre} A.,  {Toomre} J.,  1972, \mn@doi [\apj] {10.1086/151823}, \href
  {http://adsabs.harvard.edu/abs/1972ApJ...178..623T} {178, 623}

\bibitem[\protect\citeauthoryear{{Vall{\'e}e}}{{Vall{\'e}e}}{2005}]{Valle2005}
{Vall{\'e}e} J.~P.,  2005, \mn@doi [\aj] {10.1086/431744}, \href
  {http://adsabs.harvard.edu/abs/2005AJ....130..569V} {130, 569}

\bibitem[\protect\citeauthoryear{{Vall{\'e}e}}{{Vall{\'e}e}}{2008}]{Valle2008}
{Vall{\'e}e} J.~P.,  2008, \mn@doi [\aj] {10.1088/0004-6256/135/4/1301}, \href
  {http://adsabs.harvard.edu/abs/2008AJ....135.1301V} {135, 1301}

\bibitem[\protect\citeauthoryear{{Vasiliev}}{{Vasiliev}}{2019}]{agama}
{Vasiliev} E.,  2019, \mn@doi [\mnras] {10.1093/mnras/sty2672}, \href
  {https://ui.adsabs.harvard.edu/abs/2019MNRAS.482.1525V} {482, 1525}

\bibitem[\protect\citeauthoryear{{Wadsley}, {Stadel}  \& {Quinn}}{{Wadsley}
  et~al.}{2004}]{Wadsleyetal2004}
{Wadsley} J.~W.,  {Stadel} J.,   {Quinn} T.,  2004, \mn@doi [\na]
  {10.1016/j.newast.2003.08.004}, \href
  {https://ui.adsabs.harvard.edu/abs/2004NewA....9..137W} {9, 137}

\bibitem[\protect\citeauthoryear{{Wadsley}, {Keller}  \& {Quinn}}{{Wadsley}
  et~al.}{2017}]{Wadsleyetal2017}
{Wadsley} J.~W.,  {Keller} B.~W.,   {Quinn} T.~R.,  2017, \mn@doi [\mnras]
  {10.1093/mnras/stx1643}, \href
  {https://ui.adsabs.harvard.edu/abs/2017MNRAS.471.2357W} {471, 2357}

\bibitem[\protect\citeauthoryear{{Weinberg}}{{Weinberg}}{1991}]{Weinberg1991}
{Weinberg} M.~D.,  1991, \mn@doi [\apj] {10.1086/170059}, \href
  {https://ui.adsabs.harvard.edu/abs/1991ApJ...373..391W} {373, 391}

\bibitem[\protect\citeauthoryear{{Weinberg}}{{Weinberg}}{1992}]{Weinberg1992}
{Weinberg} M.~D.,  1992, \mn@doi [\apj] {10.1086/170853}, \href
  {https://ui.adsabs.harvard.edu/abs/1992ApJ...384...81W} {384, 81}

\bibitem[\protect\citeauthoryear{{Widrow}, {Gardner}, {Yanny}, {Dodelson}  \&
  {Chen}}{{Widrow} et~al.}{2012}]{Widrowetal2012}
{Widrow} L.~M.,  {Gardner} S.,  {Yanny} B.,  {Dodelson} S.,   {Chen} H.-Y.,
  2012, \mn@doi [\apjl] {10.1088/2041-8205/750/2/L41}, \href
  {https://ui.adsabs.harvard.edu/abs/2012ApJ...750L..41W} {750, L41}

\bibitem[\protect\citeauthoryear{{Widrow}, {Barber}, {Chequers}  \&
  {Cheng}}{{Widrow} et~al.}{2014}]{Widrowetal2014}
{Widrow} L.~M.,  {Barber} J.,  {Chequers} M.~H.,   {Cheng} E.,  2014, \mn@doi
  [\mnras] {10.1093/mnras/stu396}, \href
  {https://ui.adsabs.harvard.edu/abs/2014MNRAS.440.1971W} {440, 1971}

\bibitem[\protect\citeauthoryear{{Willett} et~al.,}{{Willett}
  et~al.}{2017}]{Willetetal2017}
{Willett} K.~W.,  et~al., 2017, \mn@doi [\mnras] {10.1093/mnras/stw2568}, \href
  {https://ui.adsabs.harvard.edu/abs/2017MNRAS.464.4176W} {464, 4176}

\bibitem[\protect\citeauthoryear{{Williams} et~al.,}{{Williams}
  et~al.}{2013}]{Williamsetal2013}
{Williams} M.~E.~K.,  et~al., 2013, \mn@doi [\mnras] {10.1093/mnras/stt1522},
  \href {https://ui.adsabs.harvard.edu/abs/2013MNRAS.436..101W} {436, 101}

\bibitem[\protect\citeauthoryear{{Yanny} \& {Gardner}}{{Yanny} \&
  {Gardner}}{2013}]{YannyGardner2013}
{Yanny} B.,  {Gardner} S.,  2013, \mn@doi [\apj] {10.1088/0004-637X/777/2/91},
  \href {https://ui.adsabs.harvard.edu/abs/2013ApJ...777...91Y} {777, 91}

\bibitem[\protect\citeauthoryear{{Yu}, {Ho}, {Barth}  \& {Li}}{{Yu}
  et~al.}{2018}]{Yuetal2018}
{Yu} S.-Y.,  {Ho} L.~C.,  {Barth} A.~J.,   {Li} Z.-Y.,  2018, \mn@doi [\apj]
  {10.3847/1538-4357/aacb25}, \href
  {https://ui.adsabs.harvard.edu/abs/2018ApJ...862...13Y} {862, 13}

\makeatother
\end{thebibliography}
\bibliographystyle{mnras}

%%%%%%%%%%%%%%%%%%%%%%%%%%%%%%%%%%%%%%%%%%%%%%%%%%

\bsp
\label{lastpage}

\end{document}